\documentclass[%
 reprint,
 amsmath,amssymb,
 aps,
prb,
 longbibliography
]{revtex4-2}

\usepackage{graphicx}
\usepackage{dcolumn}
\usepackage{bm}
\usepackage{physics}
\usepackage{mleftright}
\usepackage{siunitx}
\usepackage{makecell}
\usepackage{hyperref}

\usepackage{ulem}
\usepackage[usenames]{xcolor}
\usepackage{epstopdf}

\definecolor{mygreen}{RGB}{0,204,102}

\DeclareMathAlphabet{\mathpzc}{OT1}{pzc}{m}{it}

\begin{document}

\preprint{APS/123-QED}

\title{Exploring rare-earth Kitaev magnets by massive-scale computational analysis}

\author{Seong-Hoon Jang}
\affiliation{
 Institute for Materials Research, Tohoku University, 2-1-1 Katahira, Aoba-ku, Sendai, 980-8577, Japan
}
\author{Yukitoshi Motome}
\affiliation{
 Department of Applied Physics, The University of Tokyo, Tokyo 113-8656, Japan
}

\date{\today}

\begin{abstract}

The Kitaev honeycomb model plays a pivotal role in the quest for quantum spin liquids, in which fractional quasiparticles would provide applications in decoherence-free topological quantum computing.
The key ingredient is the bond-dependent Ising-type interactions, dubbed the Kitaev interactions, which require strong entanglement between spin and orbital degrees of freedom.
This study investigates the identification and design of rare-earth materials displaying robust Kitaev interactions. 
We scrutinize all possible $4f$ electron configurations, which require up to $6+$ million intermediate states in the perturbation processes, by developing a parallel computational program designed for massive scale calculations.
Our analysis reveals a predominant interplay between the isotropic Heisenberg $J$ and anisotropic Kitaev $K$ interactions across all realizations of the Kramers doublets.
Remarkably, instances featuring $4f^3$ and $4f^{11}$ configurations showcase the prevalence of $K$ over $J$, presenting unexpected prospects for exploring the Kitaev QSLs in compounds including Nd$^{3+}$ and Er$^{3+}$, respectively. 
Beyond the Kitaev model, our computational program also proves adaptable to a wide range of $4f$-electron magnets. 

\end{abstract}

\pacs{Valid PACS appear here}
\maketitle

\begin{figure*}[t!]
\includegraphics[width=2.0\columnwidth]{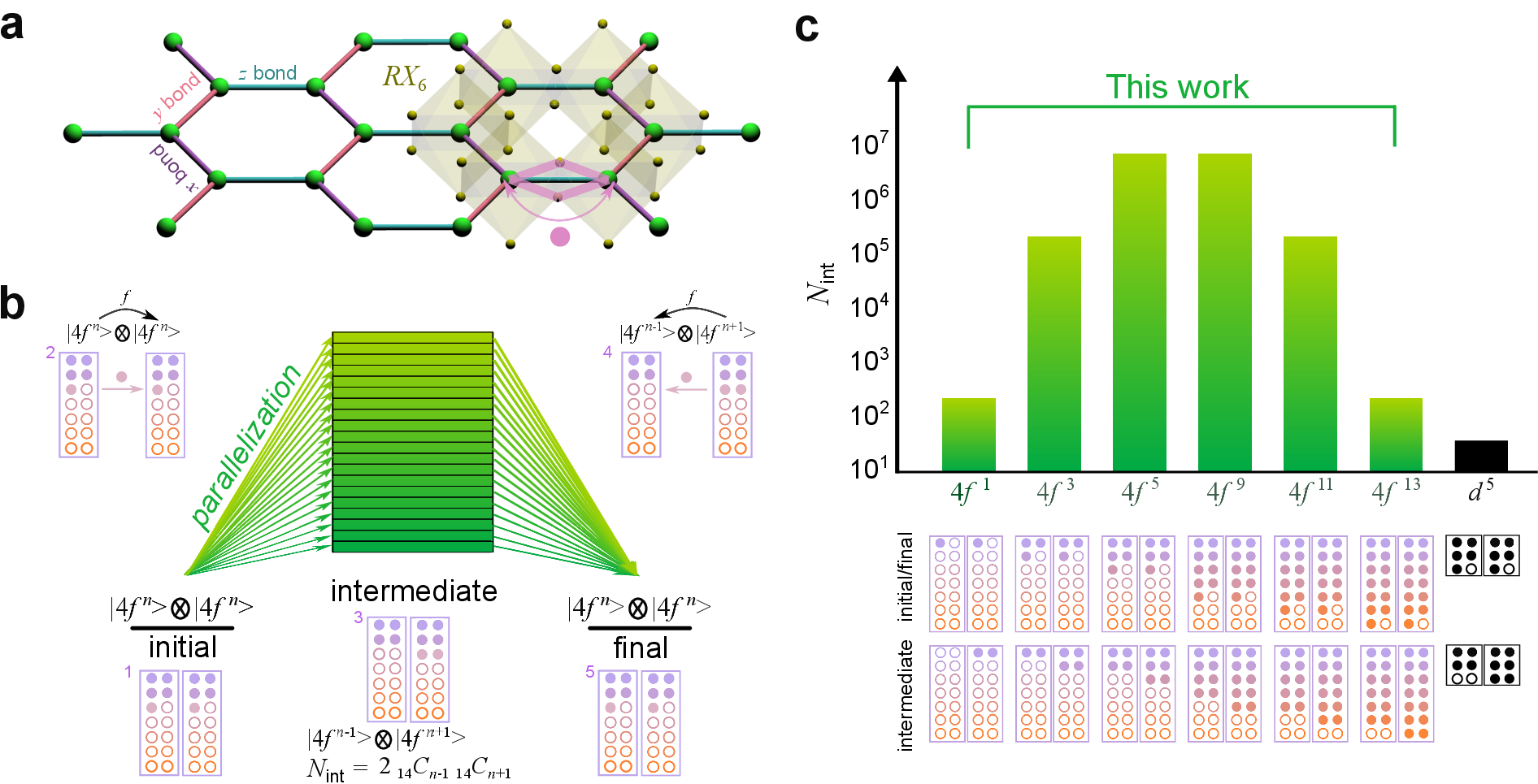}
\caption{\label{fig:scheme}
{\bf Challenge in the derivation of effective exchange interactions by second-order perturbation calculations for $4f$-electron systems.} 
{\bf a} Schematic of the Kitaev model, realized in an edge-sharing network of $RX_6$ octahedra.
Three-type $\mu$ ($x$, $y$, and $z$) bonds on the honeycomb lattice are distinguished.
The hopping paths $R$-$X$-$R$ on a $z$ bond is represented by the purple lines. 
{\bf b} Schematic of the calculations. We successfully cover the second-order perturbation calculations on a massive scale by employing a parallelization scheme spanning all the possible intermediate states $| 4f^{n-1} \rangle \otimes | 4f^{n+1} \rangle$, given the initial and final states $| 4f^n \rangle \otimes | 4f^n \rangle$.
The sequence of perturbation processes is schematically depicted for the case of $n=5$: 1 represents the initial state $|4f^5\rangle \otimes |4f^5\rangle$, 2 denotes an $f$-electron hopping from one $|4f^5\rangle$ to the other $|4f^5\rangle$, 3 represents an intermediate state $|4f^4\rangle \otimes |4f^6\rangle$, 4 denotes an $f$-electron hopping from $|4f^6\rangle$ to $|4f^4\rangle$, and 5 represents the final state $|4f^5\rangle \otimes |4f^5\rangle$ that is the same as the initial state. 
{\bf c} The number of the intermediate states, $N_{\rm int}$, for the $4f^n$-$4f^n$ states: $N_{\rm int}=182$ for $n=1$ and $n=13$, $N_{\rm int}=182,182$ for $n=3$ and $n=11$, and $N_{\rm int}=6,023,006$ for $n=5$ and $n=9$. 
The low-spin $d^5$ case with $N_{\rm int}=30$ is shown for comparison.
The initial/final and the intermediate states are schematically depicted for each electron configuration.
} 
\end{figure*}

In the realm of quantum spin liquids (QSLs), quantum fluctuations prevent localized magnetic moments from establishing conventional magnetic order, wherein excited nonlocal quasiparticles hold promise for applications in decoherence-free topological quantum computing~\cite{AN1973, SA1992, KI2006B, NA2008, BA2010, ZH2017A, SA2017}.
The Kitaev model, exhibiting exchange frustration between localized magnetic moments residing on a honeycomb lattice, is one viable model for the QSL, since it is exactly solvable through the introduction of Majorana fermions~\cite{KI2006B}.  
The Hamiltonian is given by $\mathpzc{H_{\rm Kitaev}}= \sum_{\mu}\sum_{\langle i,i' \rangle_\mu}K S_i^\mu S_{i'}^\mu$, where $K$ represents the coupling constant for the bond-dependent Ising-type interactions on three-type $\mu$ ($x$, $y$, and $z$) bonds on the honeycomb lattice, while $S_i^\mu$ ($S_{i'}^\mu$) signifies the $\mu$ component of the spin-1/2 operator at site $i$ (its nearest-neighbor site $i'$ on the $\mu$ bond).
The realization of Kitaev-type interactions has been achieved in spin-orbit coupled Mott insulators, where the interplay of electron correlation and spin-orbit coupling (SOC) is pivotal~\cite{KH2005, JA2009}. 
This is notably evident in the spin-orbital entangled Kramers doublet $\Gamma_7$, described by $j_{\rm eff}=1/2$ pseudospins, which typically originates from the low-spin $d^5$ electron configuration under the octahedral crystal field (OCF). 
Indeed, the presence of dominant Kitaev interactions has been unveiled for several quasi-two-dimensional honeycomb compounds, such as $A_2$IrO$_3$ ($A$=Na, Li), $\alpha$-RuCl$_3$, and other related materials~\cite{JA2009, TR2017, PhysRevB.93.214431, WI2017, HE2018, LI2018, SA2018, PhysRevMaterials.2.054411, KN2019, TA2019, MO2020, Motome2020, PhysRevMaterials.5.104409}. 
In these compounds, the Kitaev interactions stem from second-order perturbation processes with respect to the electron hopping mediated by ligands in edge-sharing $MX_6$ octahedra, where $M$ represents a transition metal cation and $X$ denotes a ligand ion (see Fig.~\ref{fig:scheme}{\bf a})

Besides the $d$-electron transition metal compounds, rare-earth materials with $f$ electrons meet the requirements for the Kitaev interactions: cooperation of electron correlation and SOC.  
There exist rare-earth quasi-two-dimensional honeycomb materials, which would be deemed as intriguing platforms for realizing the Kitaev model, e.g, Na$_2$PrO$_3$~\cite{HI2006}, SmI$_3$~\cite{PhysRevMaterials.6.064405}, DyCl$_3$~\cite{Templeton1954}, Er$X_3$ ($X$=Cl, Br, I)~\cite{KR1999, Kramer2000}, and YbCl$_3$~\cite{PhysRevB.102.014427}. 
Notably, the expectation is that antiferromagnetic (AFM) Kitaev interactions would manifest in the $4f^1$ electron configuration in $A_2$PrO$_3$ including Na$_2$PrO$_3$, in contrast to the ferromagnetic (FM) ones that typically dominate in $d$-electron systems~\cite{JA2019, PhysRevMaterials.4.104420, Motome2020}. 
Nonetheless, the design and discovery of $f$-electron materials with strong Kitaev-type interactions remains elusive. 
This is mainly because the formulation of low-energy effective models for $4f$-electron systems remains a formidable challenge, as it requires significant computational efforts for second-order perturbation calculations (see Fig.~\ref{fig:scheme}{\bf b}). 
Indeed, the number of the intermediate states in the perturbation processes becomes $182,182$ and $6,012,006$ for $4f^3$ ($4f^{11}$) and $4f^5$ ($4f^9$), respectively, in stark contrast to only $30$ for the low-spin $d^5$ electron configurations (see Fig.~\ref{fig:scheme}{\bf c}).

To address this issue, we develop a highly parallel computational program capable of exhaustively performing second-order perturbation calculations on a massive scale. 
The program comprises three key steps.
First, the eigenvectors $| 4f^n \rangle$ and the eigenvalues $E_{4f^n}$ for all the many-electron states with $4f^n$ electron configurations are prepared. 
In this step, the Coulomb interaction $\mathpzc{H}_{\textrm{int}}$ between $4f$ electrons is taken into account, along with the subsequent SOC $\mathpzc{H}_{\textrm{SOC}}$, based on the Russell-Saunders coupling scheme~\cite{RussellSaunders1925}.
Second, upon using these many-electron states, the initial and final $4f^n$-$4f^n$ states for neighboring sites (the tensor products $| 4f^n \rangle \otimes | 4f^n \rangle$ of $4f^n$ state pairs) and all the possible $4f^{n-1}$-$4f^{n+1}$ intermediate states (the tensor products $| 4f^{n-1} \rangle \otimes | 4f^{n+1} \rangle$ of $4f^{n-1}$ and $4f^{n+1}$ state pairs) are automatically generated.
Third, effective magnetic couplings are estimated based on the second-order perturbation expansion with respect to the $4f$ electron hopings $\mathpzc{H}_{\textrm{hop}}$. 
It is worth noting that the program can be flexibly extended beyond the second-order perturbation; it is capable of computing higher-order contributions including multiple-spin interactions.
We emphasize that even the second-order perturbation calculations are impracticable without efficient parallel computation (Fig.~\ref{fig:scheme}{\bf a}), since the number of the intermediate states exceed $6+$ million for the $4f^5$ and $4f^9$ cases (Fig.~\ref{fig:scheme}{\bf b}). 
This parallelization is achieved by implementing the Message Passing Interface in the C++ programming language.

In this study, we employ the program for the design of rare-earth Kitaev-type materials.
For the $4f^n$-$4f^n$ states with $n=1$, $3$, $5$, $9$, $11$, and $13$, we assume a perfect OCF $\mathpzc{H}_{\textrm{OCF}}$ within the edge-sharing $RX_6$ octahedra ($R$= rare-earth ions), along with $\mathpzc{H}_{\textrm{int}}$ and $\mathpzc{H}_{\textrm{SOC}}$. 
This results in the formation of spin-orbital entangled Kramers doublet for all $n$, depending on the crystal field parameters. 
In the perturbation, we take into account the indirect $4f$-$p$-$4f$ electron hoppings $\mathpzc{H}_{\textrm{hop}}$ via $p$ orbitals of ligand $X$ with the use of the Slater-Koster transfer integrals $t_{pf\pi}$ and $t_{pf\sigma}$~\cite{TA1980}, and the $p$-$4f$ energy difference $\Delta_{p\textrm{-}f}$ in the intermediate states.
Our analysis reveals that in all cases the low-energy Hamiltonian can be effectively described by two predominant exchange interactions between the pseudospins for the Kramers doublet: the bond-independent isotropic Heisenberg interaction denoted as $J$ and the bond-dependent anisotropic Kitaev interaction $K$. 
In most instances, both $J$ and $K$ exhibit AFM behavior. 
Notably, in the cases of $4f^3$ (as exemplified in Nd$^{3+}$) and the electron-hole counterpart $4f^{11}$ (Er$^{3+}$), we find that $K$ largely dominates over $J$, which realizes situations close to the pure Kitaev model. 
This finding opens up unexpected opportunities for investigating the Kitaev QSLs in $4f$-electron systems. 
Furthermore, beyond the scope of the Kitaev model, our computational program can also be applied to a wide range of $4f$-electron magnets, which would contribute to future exploration of exotic rare-earth magnetism.

\section{Results} 
\label{sec:Results}

\subsection{Ground-state Kramers doublets} 
\label{subsec:Kramers}

\begin{figure*}[th!]
\includegraphics[width=2.0\columnwidth]{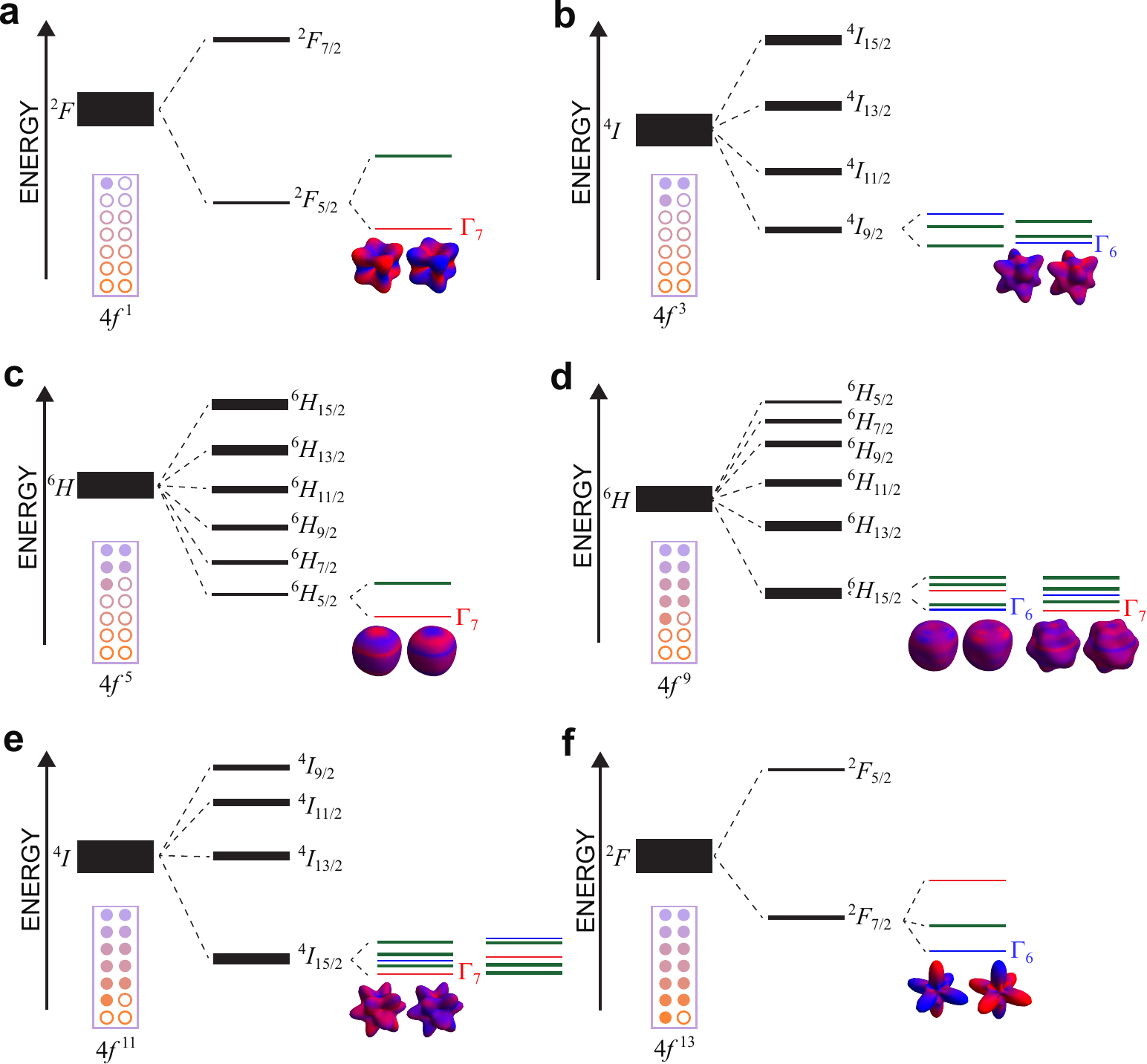}
\caption{\label{fig:Kramers}
{\bf Schematic representation of multiplet splittings for $4f^n$ electron configurations with odd integers $n$ (except for $n=7$).} 
In each configuration, the ground-state multiplet $^{2S+1}L$, initially determined by the Coulomb interaction $\mathpzc{H}_{\textrm{int}}$ (left), undergoes splitting by the spin-orbit coupling $\mathpzc{H}_{\textrm{SOC}}$, resulting in the ground-state multiplet $^{2S+1}L_J$ (middle). 
Subsequently, $^{2S+1}L_J$ is further split by the octahedral crystal field $\mathpzc{H}_{\textrm{OCF}}$, leading to the formation of ground-state Kramers doublets $\Gamma_7$ (in red) or $\Gamma_6$ (in blue), or quartets $\Gamma_6$ (in thick green). 
In the cases of $4f^3$ in {\bf b}, $4f^9$ in {\bf d}, and $4f^{11}$ in {\bf e}, the ground state are contingent upon the parameters $B_{40}$ and $B_{60}$ governing $\mathpzc{H}_{\textrm{OCF}}$; we present two extreme cases of $B_{40}=0$ (left) and $B_{60}=0$ (right).
The $4f^7$ case is not shown as the orbital is quenched.
The corresponding wave functions for the Kramers doublets are also depicted, with red and blue denoting spin-up and spin-down density profiles, respectively.
} 
\end{figure*}

\begin{widetext}
\begin{table*}[t]
\centering
\caption{\label{tab:Kramers}{\bf Ground-state multiplets and possible Kramers doublets for $4f^n$ electron configurations with odd integers $n$ (except for $n=7$).} The ground-state multiplets $^{2S+1}L_J$ are given by the Russell-Saunders coupling scheme, and the Kramers doublet $\Gamma_7$ or $\Gamma_6$ is further selected by the OCF $\mathpzc{H}_{\textrm{OCF}}$. 
In the $4f^9$ case, the ground state can be either $\Gamma_6$ or $\Gamma_7$ depending on the crystal field parameters. See also Fig.~\ref{fig:Kramers}. 
The exemplary ion is also presented for each case. 
For each Kramers doublet, the $j_{\rm eff}=1/2$ pseudospin state, where $j$ ($j^z$) is the (secondary) total angular momentum quantum number, and the coefficient $\mathbb{S}$ in Eq.~\eqref{eq:pseudospin} are explicitly shown.}
\begin{ruledtabular}
\begin{tabular}{cclc}
$4f^n$&$^{2S+1}L_J$&possible Kramers doublet&$\mathbb{S}$\\
\hline
$4f^1$ (Ce$^{3+}$)&$^2F_{5/2}$&$\Gamma_7$: $\Big| j_{\rm eff} = \pm \frac12 \Big\rangle = \frac{i}{\sqrt{6}} \bigg(-\sqrt{5} \Big| j=\frac{5}{2},j^z=\mp\frac{3}{2} \Big\rangle + \Big| j=\frac{5}{2},j^z=\pm\frac{5}{2} \Big\rangle \bigg)$ &$-3/5$\\
$4f^3$ (Nd$^{3+}$)&$^4I_{9/2}$&$\Gamma_6$: $\Big| j_{\rm eff} = \pm \frac12 \Big\rangle = \frac{i}{12} \bigg(\sqrt{6} \Big| j=\frac{9}{2},j^z=\mp\frac{7}{2} \Big\rangle + 2\sqrt{21} \Big| j=\frac{9}{2},j^z=\pm\frac{1}{2} \Big\rangle +3\sqrt{6} \Big| j=\frac{9}{2},j^z=\pm\frac{9}{2} \Big\rangle \bigg)$ &$3/11$\\
$4f^5$ (Sm$^{3+}$)&$^6H_{5/2}$&$\Gamma_7$: $\Big| j_{\rm eff} = \pm \frac12 \Big\rangle = \frac{i}{\sqrt{6}} \bigg(-\sqrt{5} \Big| j=\frac{5}{2},j^z=\mp\frac{3}{2} \Big\rangle + \Big| j=\frac{5}{2},j^z=\pm\frac{5}{2} \Big\rangle \bigg)$ &$-3/5$\\
$4f^9$ (Dy$^{3+}$)&$^6H_{15/2}$&\makecell[l]{$\Gamma_6$: $\Big| j_{\rm eff} = \pm \frac12 \Big\rangle = \frac{i}{24} \bigg(\pm\sqrt{195} \Big| j=\frac{15}{2},j^z=\mp\frac{15}{2} \Big\rangle \pm3\sqrt{7} \Big| j=\frac{15}{2},j^z=\mp\frac{7}{2} \Big\rangle$ \\ \ \ \ \ \ \ \ \ \ \ \ \ \ \ \ \ \ \ $\pm3\sqrt{33}\Big| j=\frac{15}{2},j^z=\pm\frac{1}{2} \Big\rangle \pm\sqrt{21} \Big| j=\frac{15}{2},j^z=\pm\frac{9}{2} \Big\rangle \bigg)$} &$-1/5$\\
&&\makecell[l]{$\Gamma_7$: $\Big| j_{\rm eff} = \pm \frac12 \Big\rangle = \frac{i}{24} \bigg(\pm\sqrt{33} \Big| j=\frac{15}{2},j^z=\mp\frac{11}{2} \Big\rangle 
\pm3\sqrt{13} \Big| j=\frac{15}{2},j^z=\mp\frac{3}{2} \Big\rangle \nonumber$ \\ \ \ \ \ \ \ \ \ \ \ \ \ \ \ \ \ \ \ $\mp\sqrt{195} \Big| j=\frac{15}{2},j^z=\pm\frac{5}{2} \Big\rangle \mp\sqrt{231}\Big| j=\frac{15}{2},j^z=\pm\frac{13}{2} \Big\rangle \bigg)$} &$3/17$\\
$4f^{11}$ (Er$^{3+}$)&$^4I_{15/2}$&\makecell[l]{$\Gamma_7$: $\Big| j_{\rm eff} = \pm \frac12 \Big\rangle = \frac{i}{24} \bigg(\pm\sqrt{33} \Big| j=\frac{15}{2},j^z=\mp\frac{11}{2} \Big\rangle
\pm3\sqrt{13} \Big| j=\frac{15}{2},j^z=\mp\frac{3}{2} \Big\rangle \nonumber$ \\ \ \ \ \ \ \ \ \ \ \ \ \ \ \ \ \ \ \ $\mp\sqrt{195} \Big| j=\frac{15}{2},j^z=\pm\frac{5}{2} \Big\rangle \mp\sqrt{231}\Big| j=\frac{15}{2},j^z=\pm\frac{13}{2} \Big\rangle \bigg)$} &$3/17$\\
$4f^{13}$ (Yb$^{3+}$)&$^2F_{7/2}$&$\Gamma_6$: $\Big| j_{\rm eff} = \pm \frac12 \Big\rangle = \frac{i}{6} \bigg(\mp\sqrt{15} \Big| j=\frac{7}{2},j^z=\mp\frac{7}{2} \Big\rangle \mp\sqrt{21} \Big| j=\frac{7}{2},j^z=\pm\frac{1}{2} \Big\rangle \bigg)$ &$-3/7$\\
\end{tabular}
\end{ruledtabular}
\end{table*}
\end{widetext}

Let us begin with the analysis of the crystal field splitting of the ground-state multiplets given by the Russell-Saunders coupling scheme, focusing on the $4f^n$ electron configurations with odd $n$~\cite{Motome2020}. 
In $4f^1$ electron configuration (Fig.~\ref{fig:Kramers}{\bf a}), the Coulomb interaction $\mathpzc{H}_{\textrm{int}}$ is irrelevant, leaving $14$-fold $^2F$ manifold.
This is split by $\mathpzc{H}_{\textrm{SOC}}$ into the $^2F_{5/2}$ sextet and the $^2F_{7/2}$ octet.
The ground-state $^2F_{5/2}$ sextet is further split by $\mathpzc{H}_{\textrm{OCF}}$ into the $\Gamma_7$ doublet and $\Gamma_8$ quartet. 
Since the $\Gamma_7$ doublet has lower energy than the $\Gamma_8$ quartet, the $4f^1$ case gives the $\Gamma_7$ Kramers doublet in the ground state. 
In the $4f^3$ electron configuration (Fig.~\ref{fig:Kramers}{\bf b}), $\mathpzc{H}_{\textrm{int}}$ gives $52$-fold $^4I$ manifold in the ground state, which is split by $\mathpzc{H}_{\textrm{SOC}}$ into four multiplets $^4I_{9/2}$, $^4I_{11/2}$, $^4I_{13/2}$, and $^4I_{15/2}$.
The lowest-energy $^4I_{9/2}$ dectet is further split by $\mathpzc{H}_{\textrm{OCF}}$ into the $\Gamma_6$ doublet and two $\Gamma_8$ quartets. 
The ground state depends on the crystal field parameters $B_{40}$ and $B_{60}$ (see Appendix~\ref{subsec:OCF}), and the $\Gamma_6$ Kramers doublet is selected when $B_{40}$ is predominant.
In the $4f^5$ electron configuration (Fig.~\ref{fig:Kramers}{\bf c}), the lowest-energy $^6H_{5/2}$ sextet selected by $\mathpzc{H}_{\textrm{int}}$ and $\mathpzc{H}_{\textrm{SOC}}$ is split by $\mathpzc{H}_{\textrm{OCF}}$ into the $\Gamma_7$ doublet and the $\Gamma_8$ quartet, and the ground state is given by the lower-energy $\Gamma_7$ Kramers doublet, similar to the $4f^1$ case.
In the $4f^9$ electron configuration (Fig.~\ref{fig:Kramers}{\bf d}), the lowest-energy $^6H_{15/2}$ sexdectet selected by $\mathpzc{H}_{\textrm{int}}$ and $\mathpzc{H}_{\textrm{SOC}}$ is split by $\mathpzc{H}_{\textrm{OCF}}$ into the $\Gamma_6$ doublet, the $\Gamma_7$ doublet, and the three $\Gamma_8$ quartets. 
The ground state is either the $\Gamma_6$ doublet or the $\Gamma_7$ doublet depending on the crystal field parameters.
In the $4f^{11}$ electron configuration (Fig.~\ref{fig:Kramers}{\bf e}), the ground state is given by the $\Gamma_7$ doublet when $B_{60}$ is predominant.
Finally, in the $4f^{13}$ electron configuration (Fig.~\ref{fig:Kramers}{\bf f}), the ground state is given by the $\Gamma_6$ doublet, irrespective of the crystal field parameters.
Thus, all $4f^n$ cases considered here can offer the Kramers doublet in the ground state of an isolated ion.

Table~\ref{tab:Kramers} explicitly enumerates all the accessible ground-state Kramers doublets characterized by the pseudospin $j_{\rm eff}=1/2$. 
In this table, each $j_{\rm eff}=1/2$ state is described with $j$ and $j^z$ representations; $j$ ($j^z$) is the (secondary) total angular momentum quantum number.
For the pseudospin $j_{\rm eff}=1/2$ degree of freedom, one can introduce the operator $\mathbf{S} = (S^x,S^y,S^z)^{\textrm{T}}$ defined by
\begin{equation}
S^\mu= \mathbb{S}
\begin{pmatrix} 
\mel{+}{j^\mu}{+} & \mel{+}{j^\mu}{-} \\
\mel{-}{j^\mu}{+}  & \mel{-}{j^\mu}{-}  
\end{pmatrix}
=\frac{1}{2}\sigma^\mu,
\label{eq:pseudospin}
\end{equation}
where $\mathbf{j}=(j^x,j^y,j^z)$ and $\boldsymbol{\sigma}=(\sigma^x,\sigma^y,\sigma^z)$ are the vectors of the total angular momentum operators and the Pauli matrices, respectively, and $\mathbb{S}$ is a real scalar.

\subsection{Effective exchange couplings} 
\label{subsec:couplings}

Subsequently, the program proceeds to determine the second-quantized representations with multiple $f$-orbital bases for $|4f^n\rangle \otimes |4f^n\rangle$ by the aforementioned Kramers doublets, which is commonly used for the initial and final states of the perturbation.  
Additionally, it constructs the representations for all conceivable intermediate states $|4f^{n-1}\rangle \otimes |4f^{n+1}\rangle$.
The energy difference between the initial/final states and the intermediate states is determined by two key parameters in the Hamiltonian $\mathpzc{H}_{\textrm{int}}$, namely, the onsite Coulomb interaction $U$ and the Hund's-rule coupling $J_{\rm H}$, as well as another in $\mathpzc{H}_{\textrm{SOC}}$, namely, the SOC coefficient $\lambda$. 
For obtaining the values of $U$ and $\lambda$, the Herbst-Wilkins table~\cite{HE1987} and the Freeman-Watson table~\cite{Freeman1962} are consulted, respectively.
The parameter $J_{\rm H}$ is adjusted to achieve the alignment of energy differences between different multiplets according to the Dieke diagram~\cite{Dieke1963}; see Supplementary Note~1.

\begin{figure*}[th!]
\includegraphics[width=2.0\columnwidth]{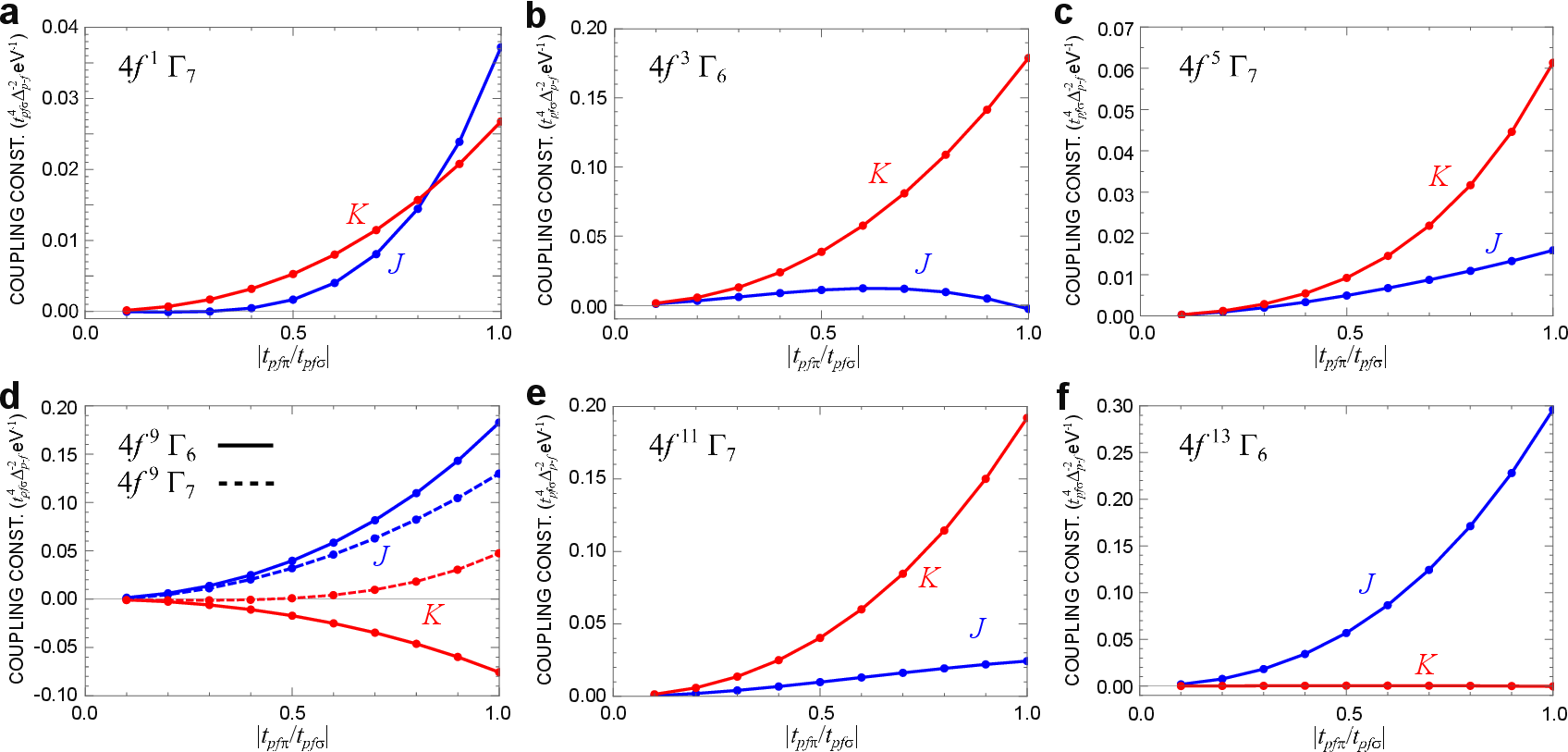}
\caption{\label{fig:JandK}
{\bf Two coupling constants, isotropic Heisenberg interaction $J$ and anisotropic Kitaev interaction $K$, derived by the second-order perturbation for $4f^n$-$4f^n$ electron configurations with odd integers $n$ (except for $n=7$).}
The data are plotted for $|t_{pf\pi}/t_{pf\sigma}|$; we take $t_{pf\pi}/t_{pf\sigma} < 0$.
} 
\end{figure*}

Given the representations and the excitation energies described above, $J$ and $K$ are calculated by employing the parallelization scheme for perturbation calculations spanning the intermediate states. 
For the hopping parameters, we adopt the Slater-Koster transfer integrals~\cite{TA1980}, while changing the ratio $|t_{pf\pi}/t_{pf\sigma}|$ between $0$ and $1$; we take $t_{pf\pi}/t_{pf\sigma} < 0$.
The results are summarized in Fig.~\ref{fig:JandK} for all $4f^n$ cases.
In the $4f^1$ $\Gamma_7$ case (Fig.~\ref{fig:JandK}{\bf a}), which includes $182$ $4f^0$-$4f^2$ intermediate states, our results emphasize the dominance of the AFM $K$ over compatibly the subdominant AFM $J$ in the wide range of $0 < |t_{pf\pi}/t_{pf\sigma}| \lesssim 0.8$. 
This behavior aligns with the findings based on the first-principles calculations in Refs.~\onlinecite{JA2019, PhysRevMaterials.4.104420}.
The magnitudes of $K$ and $J$ both monotonically increase with $|t_{pf\pi}/t_{pf\sigma}|$, which is a general trend seen also for most of the other cases below.
In the $4f^3$ $\Gamma_6$ case with $182,182$ $4f^2$-$4f^4$ intermediate states (Fig.~\ref{fig:JandK}{\bf b}), the intriguing scenario arises in which the AFM $K$ overwhelmingly outweighs non-negligible AFM $J$; this is particularly pronounced at $|t_{pf\pi}/t_{pf\sigma}| \simeq 1.0$, where $J$ almost vanishes.
We also emphasize that $K$ is one order of magnitude larger than that in the $4f^1$ case.
Note that the coupling constants $K$ and $J$ are given in unit of $t_{pf\sigma}^4 \Delta_{p-f}^{-2}$~eV$^{-1}$; we will discuss the actual values later.
In the $4f^{5}$ $\Gamma_7$ case with $6,012,006$ $4f^4$-$4f^6$ intermediate states (Fig.~\ref{fig:JandK}{\bf c}), the AFM $K$ becomes predominant compared to the subdominant AFM $J$ in the entire region of $|t_{pf\pi}/t_{pf\sigma}|$. 
In the $4f^{9}$ case (Fig.~\ref{fig:JandK}{\bf d}), there are two cases, $\Gamma_6$ and $\Gamma_7$, depending on the crystal field parameters, both of which include $6,012,006$ $4f^8$-$4f^{10}$ intermediate states. 
In the $\Gamma_6$ case, $K$ turns to be FM, while $J$ remains AFM and predominant compared to $K$.
Meanwhile, in the $\Gamma_7$ case, both $K$ and $J$ are AFM, while $J$ is again predominant.
In the case of $4f^{11}$ $\Gamma_7$ (Fig.~\ref{fig:JandK}{\bf e}), the trends mirrors the electron-hole counterpart, the $4f^3$ $\Gamma_6$ case; the AFM $K$ becomes far predominant compared to the AFM $J$, while $J$ does not decrease for large $|t_{pf\pi}/t_{pf\sigma}|$. 
We note that the magnitude of $K$ is also large comparable to the $4f^3$ case.
Finally, in the $4f^{13}$ $\Gamma_6$ case (Fig.~\ref{fig:JandK}{\bf f}), the result is quite different from the electron-hole counterpart, the $4f^1$ case; $K$ is notably suppressed compared to the predominant AFM $J$. 
This is, however, consistent with the prior findings in Ref.~\onlinecite{RA2018}.

\begin{figure}[th!]
\includegraphics[width=1.0\columnwidth]{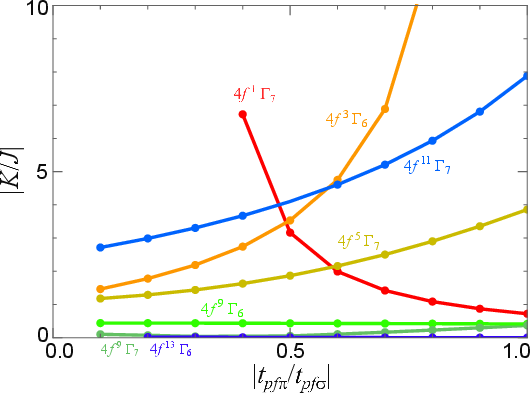}
\caption{\label{fig:KoverJ}
{\bf Ratio between the Kitaev and Heisenberg interactions, $|K/J|$, for different $4f^n$ cases.} 
We exclude the data when $|K|$ or $|J|$ are extremely small: $|K|<10^{-5}t^4_{pf\sigma}\Delta^{-2}_{p\textrm{-}f}$~eV$^{-1}$ and $|J|<10^{-5}t^4_{pf\sigma}\Delta^{-2}_{p\textrm{-}f}$~eV$^{-1}$. 
} 
\end{figure}

\begin{table}[t]
\centering
\caption{\label{tab:Values}{\bf Plausible estimates of the isotropic Heisenberg interaction $J$, anisotropic Kitaev interaction $K$, and their ratio $|K/J|$ for various Kramers doublets in $4f$-electron systems.} 
We take $t_{pf\sigma} = 0.35$~eV, $t_{pf\pi}/t_{pf\sigma} = -0.7$\cite{TA1985}, and $\Delta_{p\textrm{-}f} = 1$~eV.
}
\begin{ruledtabular}
\begin{tabular}{cccc}
 &$J$ (meV)&$K$ (meV)&$|K/J|$\\
\hline
$4f^1$ $\Gamma_7$&$0.121$&$0.172$&$1.42$\\
$4f^3$ $\Gamma_6$&$0.176$&$1.21$&$6.89$\\
$4f^5$ $\Gamma_7$&$0.131$&$0.328$&$2.50$\\
$4f^9$ $\Gamma_6$&$1.22$&$-0.521$&$0.425$\\
$4f^9$ $\Gamma_7$&$0.954$&$0.156$&$0.164$\\
$4f^{11}$ $\Gamma_7$&$0.244$&$1.27$&$5.21$\\
$4f^{13}$ $\Gamma_6$&$1.87$&$0.00566$&$0.00303$\\
\end{tabular}
\end{ruledtabular}
\end{table}

The results in Fig.~\ref{fig:JandK} highlight that, in the majority of instances, aside from the $4f^{9}$ and $4f^{13}$ cases, the AFM $K$ prevails over the subdominant AFM $J$. 
This suggests a heightened propensity for robust Kitaev interactions within diverse $4f$-electron systems. 
This is more clearly demonstrated by plotting the ratio of $|K/J|$ in Fig.~\ref{fig:KoverJ}. 
Except for the $4f^{9}$ and $4f^{13}$ cases (and the $4f^1$ case for large $|t_{pf\pi}/t_{pf\sigma}|$), $|K/J|$ is greater than $1$, indicating the predominant Kitaev interactions. 
Interestingly, besides the $4f^1$ case, $|K/J|$ consistently exhibit monotonic increases with $|t_{pf\pi}/t_{pf\sigma}|$.
It is noteworthy that the substantial predominance of AFM $K$ over AFM $J$ is particularly viable, especially in the cases of $4f^3$ $\Gamma_6$ and $4f^{11}$ $\Gamma_7$. 
In both scenarios, it is observed that $|K/J| > 4$ for $|t_{pf\pi}/t_{pf\sigma}| \gtrsim 0.6$, which includes the realistic range of the parameters~\cite{TA1985}.
In addition, the magnitude of $K$ is considerably larger than the other cases.
To emphasize these prominent properties, we show the estimates of $J$, $K$, and $|K/J|$ in Table~\ref{tab:Values}, assuming the typical values of the parameters as $t_{pf\sigma} = 0.35$~eV, $t_{pf\pi}/t_{pf\sigma} = -0.7$\cite{TA1985}, and $\Delta_{p\textrm{-}f} = 1$~eV. 
Notably, for the $4f^3$ $\Gamma_6$ and $4f^{11}$ $\Gamma_7$ configurations, it is demonstrated that $K=1.21$~meV and $1.27$~meV, respectively, which are one order of magnitude larger than the other cases, and furthermore, $K/J=6.89$ and $5.21$, signifying the substantial AFM $K$ prevalence over the AFM $J$.

\subsection{Candidate materials} 
\label{subsec:candidates}

Let us finally discuss candidate materials for the $4f^1$, $4f^3$, $4f^5$, and $4f^{11}$ cases where $K$ dominates $J$ in our calculations. 
First, for $4f^1$, the authors and their collaborators previously identified $A_2$PrO$_3$ ($A$ = alkali metals) as potential Kitaev-type magnets with the conventional assumption in the Russell-Saunders coupling scheme whereby the ordering of energy scales is given as $\mathpzc{H}_{\textrm{int}} > \mathpzc{H}_{\textrm{SOC}} \gg \mathpzc{H}_{\textrm{OCF}}$~\cite{JA2019, PhysRevMaterials.4.104420}.
However, the tetravalent Pr$^{4+}$ ion is recently recognized to reside in the intermediate coupling regime $\mathpzc{H}_{\textrm{SOC}} \sim \mathpzc{H}_{\textrm{OCF}}$~\cite{PhysRevB.103.L121109, Ramanathan2023A, Ramanathan2023B}. 
We have verified that in this regime the AFM $K$ is reduced to be subdominant, while the AFM $J$ prevails, which will be reported elsewhere.
Second, for $4f^3$, Nd$^{3+}$-based materials are considered promising, although honeycomb lattice compounds with Nd$^{3+}$ have not yet been identified to the best of our knowledge. 
This observation suggests avenues for additional materials design in the exploration of Nd$^{3+}$-based Kitaev-type magnets.
Third, for $4f^5$, SmI$_3$ has recently undergone experimental scrutiny as a potential host for the Kitaev QSL, given its absence of long-range redsout{spin} magnetic order down to $0.1$ K~\cite{PhysRevMaterials.6.064405}.
Further experiments are awaited to identify the relevant magnetic interactions.
Finally, for $4f^{11}$, Er$^{3+}$-based van der Waals magnets Er$X_3$ ($X$=Cl, Br, I) were studied~\cite{KR1999, Kramer2000}. 
These materials have similar lattice structures to that of the prime candidate for the Kitaev QSL, $\alpha$-RuCl$_3$, and were shown to exhibit noncollinear vortex-type magnetic orders.
A recent experiment for ErBr$_3$ discussed the relevance of long-range dipolar interactions~\cite{Wessler2022}.
However, note that similar vortex-like magnetic orders were also found in extensions of the Kitaev mdoel~\cite{PhysRevLett.112.077204, RU2019, PhysRevMaterials.4.104420}. 
It would be intriguing to revisit the Er$^{3+}$-based materials by using {\it ab initio} approaches.

\section{Discussion} 
\label{sec:discussion}

Our comprehensive approach, leveraging a parallel computational program capable of massive-scale second-order perturbation calculations, has provided insights into the nature of exchange interactions in rare-earth quasi-two-dimensional honeycomb lattices. 
The observed dominance of the anisotropic Kitaev interaction over the isotropic Heisenberg interaction in certain cases, particularly for $4f^3$ and $4f^{11}$ configurations, opens new avenues for investigating the Kitaev-type QSL. 
In particular, our results highlight Nd$^{3+}$ and Er$^{3+}$-based magnets as the plausible candidates for the Kitaev QSL. 
The developed computational program extends its utility beyond the Kitaev model, which would address a wide range of exchange interactions in $4f$-electron systems.
This work not only contributes to advancing our understanding of rare-earth Kitaev-type materials but also lays the groundwork for future exploration of exotic magnetism in this intriguing field of research.

\begin{acknowledgments}
The authors thank S. Bae and R. Okuma for informative discussions. 
Parts of the numerical calculations have been done using the facilities of the Supercomputer Center, the Institute for Solid State Physics, the University of Tokyo. 
This work was supported by by JST CREST (Grant No.~JP-MJCR18T2), and JSPS KAKENHI Grant Nos.~19H05825 and 20H00122.
\end{acknowledgments}

\appendix
\renewcommand\thefigure{\thesection\arabic{figure}}

\section{Methods} 
\label{sec:methods}

\subsection{Coulomb interactions} 
\label{subsec:Coulomb}

The Hamiltonian $\mathpzc{H}_{\textrm{int}}$ describing the Coulomb interactions between $f$ electrons is given by
\begin{widetext}
\begin{equation}
\mathpzc{H}_{\textrm{int}} = \sum_i \sum_{m_1,m_2,m_3,m_4} \sum_{\sigma_1,\sigma_2} \delta_{m_1+m_2,m_3+m_4} \sum_{k=0,2,4,6} F^kC^{(k)}(m_1,m_4)C^{(k)}(m_2,m_3) c^{\dagger}_{im_1\sigma_1}c^{\dagger}_{im_2\sigma_2}c_{im_3\sigma_2}c_{im_4\sigma_1},
\label{eq:H_int}
\end{equation}
\end{widetext}
where $F^k$ and $C^{(k)}$ denote the Slater-Condon parameters and the Guant coefficients, respectively ($k=0,2,4,6$); $\delta$ is the Kronecker delta; $c_{im\sigma}^\dagger$ and $c_{im\sigma}$ represent creation and annihilation operators of an electron at site $i$ in the spherical harmonics basis, respectively ($m$ and $\sigma=\pm1$ denote the magnetic and spin quantum numbers, respectively). 
Here, the Slater-Condon parameters are related with the onsite Coulomb interaction $U$ and the Hund's-rule coupling $J_{\rm H}$ as~\cite{AN1993, AN1997} 
\begin{eqnarray}
U&=&F^0,
\label{eq:U}
\\
J_{\textrm H}&=&\frac{1}{6435}\left( 286F^2+195F^4+250F^6 \right).
\label{eq:JH}
\end{eqnarray}

\subsection{Spin-orbit coupling} 
\label{subsec:SOC}

The Hamiltonian $\mathpzc{H}_{\textrm{SOC}}$ describing the effect of the SOC is given by
\begin{equation}
\mathpzc{H}_{\textrm{SOC}} = \sum_i \mathpzc{H}_{\textrm{SOC},i},
\end{equation}
where
\begin{widetext}
\begin{equation}
\mathpzc{H}_{\textrm{SOC},i} = \frac{\lambda}{2}\sum_{m=-\ell}^{\ell}\sum_{\sigma}m\sigma c^{\dagger}_{im\sigma}c_{im\sigma} + \frac{\lambda}{2}\sum_{m=-\ell}^{\ell-1}\sqrt{\ell+m+1}\sqrt{\ell-m}(c^{\dagger}_{im+1-}c_{im+}+c^{\dagger}_{im+}c_{im+1-}),
\label{eq:H_SOC}
\end{equation} 
\end{widetext}
where $\lambda>0$ is the SOC coefficient, $\ell$ is the orbital quantum number taken as $\ell=3$ for the $f$-orbital manifold. 

\subsection{Octahedral crystal field} 
\label{subsec:OCF}
The Hamiltonian $\mathpzc{H}_{\textrm{OCF}}$ describing the octahedral crystal field is given by
\begin{equation}
\mathpzc{H}_{\textrm{OCF}}= B_{40} O_4 + B_{60} O_6,
\end{equation}
where $O_4 = O_{40} + 5O_{44}$ and $O_6 = O_{60} - 21O_{64}$.
$O_{rs}$ ($s=-r, -r+1, \cdots, r$) are the rank-$r$ Stevens operators~\cite{Stevens1952}, and $B_{40}$ and $B_{60}$ are the coefficients.

\subsection{Electron hopping} 
\label{subsec:Hopping}

The Hamiltonian $\mathpzc{H}_{\textrm{hop}}$ describing the kinetic energy of electron hopping via indirect $4f$-$p$-$4f$ hopping processes is given by
\begin{equation}
\mathpzc{H}_{\textrm{hop}} = \sum_{\mu}\sum_{\langle i,i' \rangle_\mu}\mathpzc{H}^{ \left( \mu \right) }_{\textrm{hop},ii'} ,
\label{eq:Hhop} 
\end{equation}
where $\mathpzc{H}^{ \left( \mu \right) }_{\textrm{hop},ii'}$ denotes the electron hopping between nearest-neighbor sites $i$ and $i'$ on the $\mu$ bond ($\mu=x$, $y$, and $z$) as
\begin{equation}
\mathpzc{H}^{ \left( \mu \right) }_{\textrm{hop},ii'} = \sum_{m,m'}\sum_{\sigma=\pm}\left(\sum_{o,p} \frac{t_{im,op,\sigma}t_{i'm',op,\sigma}}{\Delta_{p\textrm{-}f}}c^{\dagger}_{im\sigma}c_{i'm'\sigma}+\textrm{h.c.}\right).
\label{eq:Hhopij}
\end{equation}
$t_{iu,op,\sigma}$ is the transfer integral for spin $\sigma$ between $4f$ orbital $u$ at site $i$ and $p$ orbital $p$($=x$, $y$, and $z$) at one of two ligand sites $o$($=1$ and $2$) shared by two $RX_6$ octahedra for the sites $i$ and $i'$, and $\Delta_{p\textrm{-}f}$ is the energy difference between $p$ and $4f$ orbitals.
For $t_{im,op,\sigma}$ and $t_{i'm',op,\sigma}$, we refer to Ref.~\onlinecite{TA1980}.

\subsection{Perturbation expansion}
\label{subsec:perturbation}

The effective Hamiltonian for a pair of $j_{\rm eff}=1/2$ pseudospins for nearest-neighbor sites $i$ and $i'$ on a $\mu$ bond is calculated by
\begin{widetext}
\begin{equation}
h^{\left( \mu \right)}_{ii'}=\sum_{a,b,c,d=\pm}\sum_{n}\frac{\mel{c,d}{\mathpzc{H}^{\left( \mu \right)}_{\textrm{hop},ii'}}{n}\mel{n}{\mathpzc{H}^{ \left( \mu \right) }_{\textrm{hop},ii'}}{a,b}}{E_0-E_n}\ket{c,d}\bra{a,b}.
\label{eq:h^(2)}
\end{equation}
\end{widetext}
where $\ket{a,b}$ and $\ket{c,d}$ are the initial and final two-site states with $4f^n$-$4f^n$ electron configurations described in Table~\ref{tab:Kramers} at each site, and $\ket{n}$ is the intermediate states with $4f^{n+1}$-$4f^{n-1}$ electron configurations; $E_0$ is the energy for the initial and final states, while $E_n$ is for the intermediate state $\ket{n}$. 


\bibliography{main}

\providecommand{\noopsort}[1]{}\providecommand{\singleletter}[1]{#1}%
\begin{thebibliography}{45}%
\makeatletter
\providecommand \@ifxundefined [1]{%
 \@ifx{#1\undefined}
}%
\providecommand \@ifnum [1]{%
 \ifnum #1\expandafter \@firstoftwo
 \else \expandafter \@secondoftwo
 \fi
}%
\providecommand \@ifx [1]{%
 \ifx #1\expandafter \@firstoftwo
 \else \expandafter \@secondoftwo
 \fi
}%
\providecommand \natexlab [1]{#1}%
\providecommand \enquote  [1]{``#1''}%
\providecommand \bibnamefont  [1]{#1}%
\providecommand \bibfnamefont [1]{#1}%
\providecommand \citenamefont [1]{#1}%
\providecommand \href@noop [0]{\@secondoftwo}%
\providecommand \href [0]{\begingroup \@sanitize@url \@href}%
\providecommand \@href[1]{\@@startlink{#1}\@@href}%
\providecommand \@@href[1]{\endgroup#1\@@endlink}%
\providecommand \@sanitize@url [0]{\catcode `\\12\catcode `\$12\catcode
  `\&12\catcode `\#12\catcode `\^12\catcode `\_12\catcode `\%12\relax}%
\providecommand \@@startlink[1]{}%
\providecommand \@@endlink[0]{}%
\providecommand \url  [0]{\begingroup\@sanitize@url \@url }%
\providecommand \@url [1]{\endgroup\@href {#1}{\urlprefix }}%
\providecommand \urlprefix  [0]{URL }%
\providecommand \Eprint [0]{\href }%
\providecommand \doibase [0]{https://doi.org/}%
\providecommand \selectlanguage [0]{\@gobble}%
\providecommand \bibinfo  [0]{\@secondoftwo}%
\providecommand \bibfield  [0]{\@secondoftwo}%
\providecommand \translation [1]{[#1]}%
\providecommand \BibitemOpen [0]{}%
\providecommand \bibitemStop [0]{}%
\providecommand \bibitemNoStop [0]{.\EOS\space}%
\providecommand \EOS [0]{\spacefactor3000\relax}%
\providecommand \BibitemShut  [1]{\csname bibitem#1\endcsname}%
\let\auto@bib@innerbib\@empty
\bibitem [{\citenamefont {Anderson}(1973)}]{AN1973}%
  \BibitemOpen
  \bibfield  {author} {\bibinfo {author} {\bibfnamefont {P.~W.}\ \bibnamefont
  {Anderson}},\ }\bibfield  {title} {\bibinfo {title} {Resonating valence
  bonds: A new kind of insulator?},\ }\href
  {https://doi.org/10.1016/0025-5408(73)90167-0} {\bibfield  {journal}
  {\bibinfo  {journal} {Mater. Res. Bull.}\ }\textbf {\bibinfo {volume} {8}},\
  \bibinfo {pages} {153} (\bibinfo {year} {1973})}\BibitemShut {NoStop}%
\bibitem [{\citenamefont {Sachdev}(1992)}]{SA1992}%
  \BibitemOpen
  \bibfield  {author} {\bibinfo {author} {\bibfnamefont {S.}~\bibnamefont
  {Sachdev}},\ }\bibfield  {title} {\bibinfo {title} {Kagom\'{e}- and
  triangular-lattice {H}eisenberg antiferromagnets: Ordering from quantum
  fluctuations and quantum-disordered ground states with unconfined bosonic
  spinons},\ }\href {https://doi.org/10.1103/PhysRevB.45.12377} {\bibfield
  {journal} {\bibinfo  {journal} {Phys. Rev. B}\ }\textbf {\bibinfo {volume}
  {45}},\ \bibinfo {pages} {12377} (\bibinfo {year} {1992})}\BibitemShut
  {NoStop}%
\bibitem [{\citenamefont {Kitaev}(2006)}]{KI2006B}%
  \BibitemOpen
  \bibfield  {author} {\bibinfo {author} {\bibfnamefont {A.}~\bibnamefont
  {Kitaev}},\ }\bibfield  {title} {\bibinfo {title} {Anyons in an exactly
  solved model and beyond},\ }\href
  {https://doi.org/https://doi.org/10.1016/j.aop.2005.10.005} {\bibfield
  {journal} {\bibinfo  {journal} {Ann. Phys. (N. Y.)}\ }\textbf {\bibinfo
  {volume} {321}},\ \bibinfo {pages} {2} (\bibinfo {year} {2006})}\BibitemShut
  {NoStop}%
\bibitem [{\citenamefont {Nayak}\ \emph {et~al.}(2008)\citenamefont {Nayak},
  \citenamefont {Simon}, \citenamefont {Stern}, \citenamefont {Freedman},\ and\
  \citenamefont {Das~Sarma}}]{NA2008}%
  \BibitemOpen
  \bibfield  {author} {\bibinfo {author} {\bibfnamefont {C.}~\bibnamefont
  {Nayak}}, \bibinfo {author} {\bibfnamefont {S.~H.}\ \bibnamefont {Simon}},
  \bibinfo {author} {\bibfnamefont {A.}~\bibnamefont {Stern}}, \bibinfo
  {author} {\bibfnamefont {M.}~\bibnamefont {Freedman}},\ and\ \bibinfo
  {author} {\bibfnamefont {S.}~\bibnamefont {Das~Sarma}},\ }\bibfield  {title}
  {\bibinfo {title} {Non-abelian anyons and topological quantum computation},\
  }\href {https://doi.org/10.1103/RevModPhys.80.1083} {\bibfield  {journal}
  {\bibinfo  {journal} {Rev. Mod. Phys.}\ }\textbf {\bibinfo {volume} {80}},\
  \bibinfo {pages} {1083} (\bibinfo {year} {2008})}\BibitemShut {NoStop}%
\bibitem [{\citenamefont {Balents}(2010)}]{BA2010}%
  \BibitemOpen
  \bibfield  {author} {\bibinfo {author} {\bibfnamefont {L.}~\bibnamefont
  {Balents}},\ }\bibfield  {title} {\bibinfo {title} {Spin liquids in
  frustrated magnets},\ }\href {https://doi.org/10.1038/nature08917} {\bibfield
   {journal} {\bibinfo  {journal} {Nature}\ }\textbf {\bibinfo {volume}
  {464}},\ \bibinfo {pages} {199} (\bibinfo {year} {2010})}\BibitemShut
  {NoStop}%
\bibitem [{\citenamefont {Zhou}\ \emph {et~al.}(2017)\citenamefont {Zhou},
  \citenamefont {Kanoda},\ and\ \citenamefont {Ng}}]{ZH2017A}%
  \BibitemOpen
  \bibfield  {author} {\bibinfo {author} {\bibfnamefont {Y.}~\bibnamefont
  {Zhou}}, \bibinfo {author} {\bibfnamefont {K.}~\bibnamefont {Kanoda}},\ and\
  \bibinfo {author} {\bibfnamefont {T.-K.}\ \bibnamefont {Ng}},\ }\bibfield
  {title} {\bibinfo {title} {Quantum spin liquid states},\ }\href
  {https://doi.org/10.1103/RevModPhys.89.025003} {\bibfield  {journal}
  {\bibinfo  {journal} {Rev. Mod. Phys.}\ }\textbf {\bibinfo {volume} {89}},\
  \bibinfo {pages} {025003} (\bibinfo {year} {2017})}\BibitemShut {NoStop}%
\bibitem [{\citenamefont {Savary}\ and\ \citenamefont
  {Balents}(2016)}]{SA2017}%
  \BibitemOpen
  \bibfield  {author} {\bibinfo {author} {\bibfnamefont {L.}~\bibnamefont
  {Savary}}\ and\ \bibinfo {author} {\bibfnamefont {L.}~\bibnamefont
  {Balents}},\ }\bibfield  {title} {\bibinfo {title} {Quantum spin liquids: a
  review},\ }\href {https://doi.org/10.1088/0034-4885/80/1/016502} {\bibfield
  {journal} {\bibinfo  {journal} {Rep. Prog. Phys.}\ }\textbf {\bibinfo
  {volume} {80}},\ \bibinfo {pages} {016502} (\bibinfo {year}
  {2016})}\BibitemShut {NoStop}%
\bibitem [{\citenamefont {Khaliullin}(2005)}]{KH2005}%
  \BibitemOpen
  \bibfield  {author} {\bibinfo {author} {\bibfnamefont {G.}~\bibnamefont
  {Khaliullin}},\ }\bibfield  {title} {\bibinfo {title} {Orbital order and
  fluctuations in {M}ott insulators},\ }\href
  {https://doi.org/10.1143/PTPS.160.155} {\bibfield  {journal} {\bibinfo
  {journal} {Prog. Theor. Phys. Suppl.}\ }\textbf {\bibinfo {volume} {160}},\
  \bibinfo {pages} {155} (\bibinfo {year} {2005})}\BibitemShut {NoStop}%
\bibitem [{\citenamefont {Jackeli}\ and\ \citenamefont
  {Khaliullin}(2009)}]{JA2009}%
  \BibitemOpen
  \bibfield  {author} {\bibinfo {author} {\bibfnamefont {G.}~\bibnamefont
  {Jackeli}}\ and\ \bibinfo {author} {\bibfnamefont {G.}~\bibnamefont
  {Khaliullin}},\ }\bibfield  {title} {\bibinfo {title} {Mott insulators in the
  strong spin-orbit coupling limit: From {H}eisenberg to a quantum compass and
  {K}itaev models},\ }\href {https://doi.org/10.1103/PhysRevLett.102.017205}
  {\bibfield  {journal} {\bibinfo  {journal} {Phys. Rev. Lett.}\ }\textbf
  {\bibinfo {volume} {102}},\ \bibinfo {pages} {017205} (\bibinfo {year}
  {2009})}\BibitemShut {NoStop}%
\bibitem [{\citenamefont {Trebst}(2017)}]{TR2017}%
  \BibitemOpen
  \bibfield  {author} {\bibinfo {author} {\bibfnamefont {S.}~\bibnamefont
  {Trebst}},\ }\bibfield  {title} {\bibinfo {title} {{Kitaev materials}},\
  }\Eprint {https://arxiv.org/abs/cond-mat.str-el/1701.07056}
  {arXiv:cond-mat.str-el/1701.07056}  (\bibinfo {year} {2017})\BibitemShut
  {NoStop}%
\bibitem [{\citenamefont {Winter}\ \emph {et~al.}(2016)\citenamefont {Winter},
  \citenamefont {Li}, \citenamefont {Jeschke},\ and\ \citenamefont
  {Valent\'{\i}}}]{PhysRevB.93.214431}%
  \BibitemOpen
  \bibfield  {author} {\bibinfo {author} {\bibfnamefont {S.~M.}\ \bibnamefont
  {Winter}}, \bibinfo {author} {\bibfnamefont {Y.}~\bibnamefont {Li}}, \bibinfo
  {author} {\bibfnamefont {H.~O.}\ \bibnamefont {Jeschke}},\ and\ \bibinfo
  {author} {\bibfnamefont {R.}~\bibnamefont {Valent\'{\i}}},\ }\bibfield
  {title} {\bibinfo {title} {Challenges in design of {K}itaev materials:
  Magnetic interactions from competing energy scales},\ }\href
  {https://doi.org/10.1103/PhysRevB.93.214431} {\bibfield  {journal} {\bibinfo
  {journal} {Phys. Rev. B}\ }\textbf {\bibinfo {volume} {93}},\ \bibinfo
  {pages} {214431} (\bibinfo {year} {2016})}\BibitemShut {NoStop}%
\bibitem [{\citenamefont {Winter}\ \emph {et~al.}(2017)\citenamefont {Winter},
  \citenamefont {Tsirlin}, \citenamefont {Daghofer}, \citenamefont {van~den
  Brink}, \citenamefont {Singh}, \citenamefont {Gegenwart},\ and\ \citenamefont
  {Valent{\'{\i}}}}]{WI2017}%
  \BibitemOpen
  \bibfield  {author} {\bibinfo {author} {\bibfnamefont {S.~M.}\ \bibnamefont
  {Winter}}, \bibinfo {author} {\bibfnamefont {A.~A.}\ \bibnamefont {Tsirlin}},
  \bibinfo {author} {\bibfnamefont {M.}~\bibnamefont {Daghofer}}, \bibinfo
  {author} {\bibfnamefont {J.}~\bibnamefont {van~den Brink}}, \bibinfo {author}
  {\bibfnamefont {Y.}~\bibnamefont {Singh}}, \bibinfo {author} {\bibfnamefont
  {P.}~\bibnamefont {Gegenwart}},\ and\ \bibinfo {author} {\bibfnamefont
  {R.}~\bibnamefont {Valent{\'{\i}}}},\ }\bibfield  {title} {\bibinfo {title}
  {Models and materials for generalized {K}itaev magnetism},\ }\href
  {https://doi.org/10.1088/1361-648x/aa8cf5} {\bibfield  {journal} {\bibinfo
  {journal} {J. Phys.: Condens. Matter}\ }\textbf {\bibinfo {volume} {29}},\
  \bibinfo {pages} {493002} (\bibinfo {year} {2017})}\BibitemShut {NoStop}%
\bibitem [{\citenamefont {Hermanns}\ \emph {et~al.}(2018)\citenamefont
  {Hermanns}, \citenamefont {Kimchi},\ and\ \citenamefont {Knolle}}]{HE2018}%
  \BibitemOpen
  \bibfield  {author} {\bibinfo {author} {\bibfnamefont {M.}~\bibnamefont
  {Hermanns}}, \bibinfo {author} {\bibfnamefont {I.}~\bibnamefont {Kimchi}},\
  and\ \bibinfo {author} {\bibfnamefont {J.}~\bibnamefont {Knolle}},\
  }\bibfield  {title} {\bibinfo {title} {Physics of the {K}itaev model:
  Fractionalization, dynamic correlations, and material connections},\ }\href
  {https://doi.org/10.1146/annurev-conmatphys-033117-053934} {\bibfield
  {journal} {\bibinfo  {journal} {Annu. Rev. Condens. Matter Phys.}\ }\textbf
  {\bibinfo {volume} {9}},\ \bibinfo {pages} {17} (\bibinfo {year}
  {2018})}\BibitemShut {NoStop}%
\bibitem [{\citenamefont {Liu}\ and\ \citenamefont
  {Khaliullin}(2018)}]{LI2018}%
  \BibitemOpen
  \bibfield  {author} {\bibinfo {author} {\bibfnamefont {H.}~\bibnamefont
  {Liu}}\ and\ \bibinfo {author} {\bibfnamefont {G.}~\bibnamefont
  {Khaliullin}},\ }\bibfield  {title} {\bibinfo {title} {Pseudospin exchange
  interactions in ${d}^{7}$ cobalt compounds: Possible realization of the
  {K}itaev model},\ }\href {https://doi.org/10.1103/PhysRevB.97.014407}
  {\bibfield  {journal} {\bibinfo  {journal} {Phys. Rev. B}\ }\textbf {\bibinfo
  {volume} {97}},\ \bibinfo {pages} {014407} (\bibinfo {year}
  {2018})}\BibitemShut {NoStop}%
\bibitem [{\citenamefont {Sano}\ \emph {et~al.}(2018)\citenamefont {Sano},
  \citenamefont {Kato},\ and\ \citenamefont {Motome}}]{SA2018}%
  \BibitemOpen
  \bibfield  {author} {\bibinfo {author} {\bibfnamefont {R.}~\bibnamefont
  {Sano}}, \bibinfo {author} {\bibfnamefont {Y.}~\bibnamefont {Kato}},\ and\
  \bibinfo {author} {\bibfnamefont {Y.}~\bibnamefont {Motome}},\ }\bibfield
  {title} {\bibinfo {title} {{K}itaev-{H}eisenberg hamiltonian for high-spin
  ${d}^{7}$ {M}ott insulators},\ }\href
  {https://doi.org/10.1103/PhysRevB.97.014408} {\bibfield  {journal} {\bibinfo
  {journal} {Phys. Rev. B}\ }\textbf {\bibinfo {volume} {97}},\ \bibinfo
  {pages} {014408} (\bibinfo {year} {2018})}\BibitemShut {NoStop}%
\bibitem [{\citenamefont {Haraguchi}\ \emph {et~al.}(2018)\citenamefont
  {Haraguchi}, \citenamefont {Michioka}, \citenamefont {Matsuo}, \citenamefont
  {Kindo}, \citenamefont {Ueda},\ and\ \citenamefont
  {Yoshimura}}]{PhysRevMaterials.2.054411}%
  \BibitemOpen
  \bibfield  {author} {\bibinfo {author} {\bibfnamefont {Y.}~\bibnamefont
  {Haraguchi}}, \bibinfo {author} {\bibfnamefont {C.}~\bibnamefont {Michioka}},
  \bibinfo {author} {\bibfnamefont {A.}~\bibnamefont {Matsuo}}, \bibinfo
  {author} {\bibfnamefont {K.}~\bibnamefont {Kindo}}, \bibinfo {author}
  {\bibfnamefont {H.}~\bibnamefont {Ueda}},\ and\ \bibinfo {author}
  {\bibfnamefont {K.}~\bibnamefont {Yoshimura}},\ }\bibfield  {title} {\bibinfo
  {title} {Magnetic ordering with an ${X}$${Y}$-like anisotropy in the
  honeycomb lattice iridates {Z}n{I}r{O}$_3$ and {M}g{I}r{O}$_3$ synthesized
  via a metathesis reaction},\ }\href
  {https://doi.org/10.1103/PhysRevMaterials.2.054411} {\bibfield  {journal}
  {\bibinfo  {journal} {Phys. Rev. Mater.}\ }\textbf {\bibinfo {volume} {2}},\
  \bibinfo {pages} {054411} (\bibinfo {year} {2018})}\BibitemShut {NoStop}%
\bibitem [{\citenamefont {Knolle}\ and\ \citenamefont
  {Moessner}(2019)}]{KN2019}%
  \BibitemOpen
  \bibfield  {author} {\bibinfo {author} {\bibfnamefont {J.}~\bibnamefont
  {Knolle}}\ and\ \bibinfo {author} {\bibfnamefont {R.}~\bibnamefont
  {Moessner}},\ }\bibfield  {title} {\bibinfo {title} {A field guide to spin
  liquids},\ }\href {https://doi.org/10.1146/annurev-conmatphys-031218-013401}
  {\bibfield  {journal} {\bibinfo  {journal} {Annu. Rev. Condens. Matter
  Phys.}\ }\textbf {\bibinfo {volume} {10}},\ \bibinfo {pages} {451} (\bibinfo
  {year} {2019})}\BibitemShut {NoStop}%
\bibitem [{\citenamefont {Takagi}\ \emph {et~al.}(2019)\citenamefont {Takagi},
  \citenamefont {Takayama}, \citenamefont {Jackeli}, \citenamefont
  {Khaliullin},\ and\ \citenamefont {Nagler}}]{TA2019}%
  \BibitemOpen
  \bibfield  {author} {\bibinfo {author} {\bibfnamefont {H.}~\bibnamefont
  {Takagi}}, \bibinfo {author} {\bibfnamefont {T.}~\bibnamefont {Takayama}},
  \bibinfo {author} {\bibfnamefont {G.}~\bibnamefont {Jackeli}}, \bibinfo
  {author} {\bibfnamefont {G.}~\bibnamefont {Khaliullin}},\ and\ \bibinfo
  {author} {\bibfnamefont {S.~E.}\ \bibnamefont {Nagler}},\ }\bibfield  {title}
  {\bibinfo {title} {Concept and realization of {K}itaev quantum spin
  liquids},\ }\href {https://doi.org/10.1038/s42254-019-0038-2} {\bibfield
  {journal} {\bibinfo  {journal} {Nat. Rev. Phys.}\ }\textbf {\bibinfo {volume}
  {1}},\ \bibinfo {pages} {264} (\bibinfo {year} {2019})}\BibitemShut {NoStop}%
\bibitem [{\citenamefont {Motome}\ and\ \citenamefont {Nasu}(2020)}]{MO2020}%
  \BibitemOpen
  \bibfield  {author} {\bibinfo {author} {\bibfnamefont {Y.}~\bibnamefont
  {Motome}}\ and\ \bibinfo {author} {\bibfnamefont {J.}~\bibnamefont {Nasu}},\
  }\bibfield  {title} {\bibinfo {title} {{Hunting Majorana Fermions in Kitaev
  magnets}},\ }\href {https://doi.org/10.7566/JPSJ.89.012002} {\bibfield
  {journal} {\bibinfo  {journal} {J. Phys. Soc. Jpn.}\ }\textbf {\bibinfo
  {volume} {89}},\ \bibinfo {pages} {012002} (\bibinfo {year}
  {2020})}\BibitemShut {NoStop}%
\bibitem [{\citenamefont {Motome}\ \emph {et~al.}(2020)\citenamefont {Motome},
  \citenamefont {Sano}, \citenamefont {Jang}, \citenamefont {Sugita},\ and\
  \citenamefont {Kato}}]{Motome2020}%
  \BibitemOpen
  \bibfield  {author} {\bibinfo {author} {\bibfnamefont {Y.}~\bibnamefont
  {Motome}}, \bibinfo {author} {\bibfnamefont {R.}~\bibnamefont {Sano}},
  \bibinfo {author} {\bibfnamefont {S.-H.}\ \bibnamefont {Jang}}, \bibinfo
  {author} {\bibfnamefont {Y.}~\bibnamefont {Sugita}},\ and\ \bibinfo {author}
  {\bibfnamefont {Y.}~\bibnamefont {Kato}},\ }\bibfield  {title} {\bibinfo
  {title} {Materials design of {K}itaev spin liquids beyond the
  {J}ackeli–{K}haliullin mechanism},\ }\href
  {https://doi.org/10.1088/1361-648X/ab8525} {\bibfield  {journal} {\bibinfo
  {journal} {J. Phys.: Condens. Matter}\ }\textbf {\bibinfo {volume} {32}},\
  \bibinfo {pages} {404001} (\bibinfo {year} {2020})}\BibitemShut {NoStop}%
\bibitem [{\citenamefont {Jang}\ and\ \citenamefont
  {Motome}(2021)}]{PhysRevMaterials.5.104409}%
  \BibitemOpen
  \bibfield  {author} {\bibinfo {author} {\bibfnamefont {S.-H.}\ \bibnamefont
  {Jang}}\ and\ \bibinfo {author} {\bibfnamefont {Y.}~\bibnamefont {Motome}},\
  }\bibfield  {title} {\bibinfo {title} {Electronic and magnetic properties of
  iridium ilmenites ${A}${I}r{O}$_3$ (${A}=$ {M}g, {Z}n, and {M}n)},\ }\href
  {https://doi.org/10.1103/PhysRevMaterials.5.104409} {\bibfield  {journal}
  {\bibinfo  {journal} {Phys. Rev. Mater.}\ }\textbf {\bibinfo {volume} {5}},\
  \bibinfo {pages} {104409} (\bibinfo {year} {2021})}\BibitemShut {NoStop}%
\bibitem [{\citenamefont {Hinatsu}\ and\ \citenamefont {Doi}(2006)}]{HI2006}%
  \BibitemOpen
  \bibfield  {author} {\bibinfo {author} {\bibfnamefont {Y.}~\bibnamefont
  {Hinatsu}}\ and\ \bibinfo {author} {\bibfnamefont {Y.}~\bibnamefont {Doi}},\
  }\bibfield  {title} {\bibinfo {title} {Crystal structures and magnetic
  properties of alkali-metal lanthanide oxides ${A}$$_2${L}n{O}$_3$
  (${A}$={L}i, {N}a; {L}n={C}e, {P}r, {T}b)},\ }\href
  {https://doi.org/https://doi.org/10.1016/j.jallcom.2005.08.100} {\bibfield
  {journal} {\bibinfo  {journal} {J. Alloy. Comp.}\ }\textbf {\bibinfo {volume}
  {418}},\ \bibinfo {pages} {155} (\bibinfo {year} {2006})}\BibitemShut
  {NoStop}%
\bibitem [{\citenamefont {Ishikawa}\ \emph {et~al.}(2022)\citenamefont
  {Ishikawa}, \citenamefont {Kurihara}, \citenamefont {Yajima}, \citenamefont
  {Nishio-Hamane}, \citenamefont {Shimizu}, \citenamefont {Sakakibara},
  \citenamefont {Matsuo},\ and\ \citenamefont
  {Kindo}}]{PhysRevMaterials.6.064405}%
  \BibitemOpen
  \bibfield  {author} {\bibinfo {author} {\bibfnamefont {H.}~\bibnamefont
  {Ishikawa}}, \bibinfo {author} {\bibfnamefont {R.}~\bibnamefont {Kurihara}},
  \bibinfo {author} {\bibfnamefont {T.}~\bibnamefont {Yajima}}, \bibinfo
  {author} {\bibfnamefont {D.}~\bibnamefont {Nishio-Hamane}}, \bibinfo {author}
  {\bibfnamefont {Y.}~\bibnamefont {Shimizu}}, \bibinfo {author} {\bibfnamefont
  {T.}~\bibnamefont {Sakakibara}}, \bibinfo {author} {\bibfnamefont
  {A.}~\bibnamefont {Matsuo}},\ and\ \bibinfo {author} {\bibfnamefont
  {K.}~\bibnamefont {Kindo}},\ }\bibfield  {title} {\bibinfo {title}
  {{S}m{I}$_3$: $4f^5$ honeycomb magnet with spin-orbital entangled
  ${\mathrm{\ensuremath{\Gamma}}}_{7}$ kramers doublet},\ }\href
  {https://doi.org/10.1103/PhysRevMaterials.6.064405} {\bibfield  {journal}
  {\bibinfo  {journal} {Phys. Rev. Mater.}\ }\textbf {\bibinfo {volume} {6}},\
  \bibinfo {pages} {064405} (\bibinfo {year} {2022})}\BibitemShut {NoStop}%
\bibitem [{\citenamefont {Templeton}\ and\ \citenamefont
  {Carter}(1954)}]{Templeton1954}%
  \BibitemOpen
  \bibfield  {author} {\bibinfo {author} {\bibfnamefont {D.~H.}\ \bibnamefont
  {Templeton}}\ and\ \bibinfo {author} {\bibfnamefont {G.~F.}\ \bibnamefont
  {Carter}},\ }\bibfield  {title} {\bibinfo {title} {The crystal structures of
  yttrium trichloride and similar compounds},\ }\href
  {https://doi.org/10.1021/j150521a002} {\bibfield  {journal} {\bibinfo
  {journal} {J. Phys. Chem.}\ }\textbf {\bibinfo {volume} {58}},\ \bibinfo
  {pages} {940} (\bibinfo {year} {1954})}\BibitemShut {NoStop}%
\bibitem [{\citenamefont {Kr\"amer}\ \emph {et~al.}(1999)\citenamefont
  {Kr\"amer}, \citenamefont {G\"udel}, \citenamefont {Roessli}, \citenamefont
  {Fischer}, \citenamefont {D\"onni}, \citenamefont {Wada}, \citenamefont
  {Fauth}, \citenamefont {Fernandez-Diaz},\ and\ \citenamefont
  {Hauss}}]{KR1999}%
  \BibitemOpen
  \bibfield  {author} {\bibinfo {author} {\bibfnamefont {K.~W.}\ \bibnamefont
  {Kr\"amer}}, \bibinfo {author} {\bibfnamefont {H.~U.}\ \bibnamefont
  {G\"udel}}, \bibinfo {author} {\bibfnamefont {B.}~\bibnamefont {Roessli}},
  \bibinfo {author} {\bibfnamefont {P.}~\bibnamefont {Fischer}}, \bibinfo
  {author} {\bibfnamefont {A.}~\bibnamefont {D\"onni}}, \bibinfo {author}
  {\bibfnamefont {N.}~\bibnamefont {Wada}}, \bibinfo {author} {\bibfnamefont
  {F.}~\bibnamefont {Fauth}}, \bibinfo {author} {\bibfnamefont {M.~T.}\
  \bibnamefont {Fernandez-Diaz}},\ and\ \bibinfo {author} {\bibfnamefont
  {T.}~\bibnamefont {Hauss}},\ }\bibfield  {title} {\bibinfo {title}
  {Noncollinear two- and three-dimensional magnetic ordering in the honeycomb
  lattices of {E}r${X}_3$ (${X}$ = {C}l, {B}r, {I})},\ }\href
  {https://doi.org/10.1103/PhysRevB.60.R3724} {\bibfield  {journal} {\bibinfo
  {journal} {Phys. Rev. B}\ }\textbf {\bibinfo {volume} {60}},\ \bibinfo
  {pages} {R3724} (\bibinfo {year} {1999})}\BibitemShut {NoStop}%
\bibitem [{\citenamefont {Kr\"amer}\ \emph {et~al.}(2000)\citenamefont
  {Kr\"amer}, \citenamefont {G\"udel}, \citenamefont {Fischer}, \citenamefont
  {Fauth}, \citenamefont {Fernandez-Diaz},\ and\ \citenamefont
  {Hau\ss{}}}]{Kramer2000}%
  \BibitemOpen
  \bibfield  {author} {\bibinfo {author} {\bibfnamefont {K.~W.}\ \bibnamefont
  {Kr\"amer}}, \bibinfo {author} {\bibfnamefont {H.~U.}\ \bibnamefont
  {G\"udel}}, \bibinfo {author} {\bibfnamefont {P.}~\bibnamefont {Fischer}},
  \bibinfo {author} {\bibfnamefont {F.}~\bibnamefont {Fauth}}, \bibinfo
  {author} {\bibfnamefont {M.~T.}\ \bibnamefont {Fernandez-Diaz}},\ and\
  \bibinfo {author} {\bibfnamefont {T.}~\bibnamefont {Hau\ss{}}},\ }\bibfield
  {title} {\bibinfo {title} {Triangular antiferromagnetic order in the
  honeycomb layer lattice of {E}r{C}l$_3$},\ }\href
  {https://doi.org/10.1007/s100510070075} {\bibfield  {journal} {\bibinfo
  {journal} {Eur. Phys. J. B}\ }\textbf {\bibinfo {volume} {18}},\ \bibinfo
  {pages} {39} (\bibinfo {year} {2000})}\BibitemShut {NoStop}%
\bibitem [{\citenamefont {Xing}\ \emph {et~al.}(2020)\citenamefont {Xing},
  \citenamefont {Feng}, \citenamefont {Liu}, \citenamefont {Emmanouilidou},
  \citenamefont {Hu}, \citenamefont {Liu}, \citenamefont {Graf}, \citenamefont
  {Ramirez}, \citenamefont {Chen}, \citenamefont {Cao},\ and\ \citenamefont
  {Ni}}]{PhysRevB.102.014427}%
  \BibitemOpen
  \bibfield  {author} {\bibinfo {author} {\bibfnamefont {J.}~\bibnamefont
  {Xing}}, \bibinfo {author} {\bibfnamefont {E.}~\bibnamefont {Feng}}, \bibinfo
  {author} {\bibfnamefont {Y.}~\bibnamefont {Liu}}, \bibinfo {author}
  {\bibfnamefont {E.}~\bibnamefont {Emmanouilidou}}, \bibinfo {author}
  {\bibfnamefont {C.}~\bibnamefont {Hu}}, \bibinfo {author} {\bibfnamefont
  {J.}~\bibnamefont {Liu}}, \bibinfo {author} {\bibfnamefont {D.}~\bibnamefont
  {Graf}}, \bibinfo {author} {\bibfnamefont {A.~P.}\ \bibnamefont {Ramirez}},
  \bibinfo {author} {\bibfnamefont {G.}~\bibnamefont {Chen}}, \bibinfo {author}
  {\bibfnamefont {H.}~\bibnamefont {Cao}},\ and\ \bibinfo {author}
  {\bibfnamefont {N.}~\bibnamefont {Ni}},\ }\bibfield  {title} {\bibinfo
  {title} {N\'eel-type antiferromagnetic order and magnetic field--temperature
  phase diagram in the spin-$\frac{1}{2}$ rare-earth honeycomb compound
  $\mathrm{YbCl}{}_{3}$},\ }\href {https://doi.org/10.1103/PhysRevB.102.014427}
  {\bibfield  {journal} {\bibinfo  {journal} {Phys. Rev. B}\ }\textbf {\bibinfo
  {volume} {102}},\ \bibinfo {pages} {014427} (\bibinfo {year}
  {2020})}\BibitemShut {NoStop}%
\bibitem [{\citenamefont {Jang}\ \emph {et~al.}(2019)\citenamefont {Jang},
  \citenamefont {Sano}, \citenamefont {Kato},\ and\ \citenamefont
  {Motome}}]{JA2019}%
  \BibitemOpen
  \bibfield  {author} {\bibinfo {author} {\bibfnamefont {S.-H.}\ \bibnamefont
  {Jang}}, \bibinfo {author} {\bibfnamefont {R.}~\bibnamefont {Sano}}, \bibinfo
  {author} {\bibfnamefont {Y.}~\bibnamefont {Kato}},\ and\ \bibinfo {author}
  {\bibfnamefont {Y.}~\bibnamefont {Motome}},\ }\bibfield  {title} {\bibinfo
  {title} {Antiferromagnetic {K}itaev interaction in $f$-electron based
  honeycomb magnets},\ }\href {https://doi.org/10.1103/PhysRevB.99.241106}
  {\bibfield  {journal} {\bibinfo  {journal} {Phys. Rev. B}\ }\textbf {\bibinfo
  {volume} {99}},\ \bibinfo {pages} {241106(R)} (\bibinfo {year}
  {2019})}\BibitemShut {NoStop}%
\bibitem [{\citenamefont {Jang}\ \emph {et~al.}(2020)\citenamefont {Jang},
  \citenamefont {Sano}, \citenamefont {Kato},\ and\ \citenamefont
  {Motome}}]{PhysRevMaterials.4.104420}%
  \BibitemOpen
  \bibfield  {author} {\bibinfo {author} {\bibfnamefont {S.-H.}\ \bibnamefont
  {Jang}}, \bibinfo {author} {\bibfnamefont {R.}~\bibnamefont {Sano}}, \bibinfo
  {author} {\bibfnamefont {Y.}~\bibnamefont {Kato}},\ and\ \bibinfo {author}
  {\bibfnamefont {Y.}~\bibnamefont {Motome}},\ }\bibfield  {title} {\bibinfo
  {title} {Computational design of $f$-electron kitaev magnets: Honeycomb and
  hyperhoneycomb compounds ${A}_2${P}r{O}$_3$ (${A}=$ alkali metals)},\ }\href
  {https://doi.org/10.1103/PhysRevMaterials.4.104420} {\bibfield  {journal}
  {\bibinfo  {journal} {Phys. Rev. Mater.}\ }\textbf {\bibinfo {volume} {4}},\
  \bibinfo {pages} {104420} (\bibinfo {year} {2020})}\BibitemShut {NoStop}%
\bibitem [{\citenamefont {Russell}\ and\ \citenamefont
  {Saunders}(1925)}]{RussellSaunders1925}%
  \BibitemOpen
  \bibfield  {author} {\bibinfo {author} {\bibfnamefont {H.~N.}\ \bibnamefont
  {Russell}}\ and\ \bibinfo {author} {\bibfnamefont {F.~A.}\ \bibnamefont
  {Saunders}},\ }\bibfield  {title} {\bibinfo {title} {New regularities in the
  spectra of the alkaline earths},\ }\href {https://doi.org/10.1086/142872}
  {\bibfield  {journal} {\bibinfo  {journal} {Astrophysical Journal}\ }\textbf
  {\bibinfo {volume} {61}},\ \bibinfo {pages} {38} (\bibinfo {year}
  {1925})}\BibitemShut {NoStop}%
\bibitem [{\citenamefont {Takegahara}\ \emph {et~al.}(1980)\citenamefont
  {Takegahara}, \citenamefont {Aoki},\ and\ \citenamefont {Yanase}}]{TA1980}%
  \BibitemOpen
  \bibfield  {author} {\bibinfo {author} {\bibfnamefont {K.}~\bibnamefont
  {Takegahara}}, \bibinfo {author} {\bibfnamefont {Y.}~\bibnamefont {Aoki}},\
  and\ \bibinfo {author} {\bibfnamefont {A.}~\bibnamefont {Yanase}},\
  }\bibfield  {title} {\bibinfo {title} {Slater-{K}oster tables for $f$
  electrons},\ }\href {https://doi.org/10.1088/0022-3719/13/4/016} {\bibfield
  {journal} {\bibinfo  {journal} {J. Phys. C: Solid St. Phys.}\ }\textbf
  {\bibinfo {volume} {13}},\ \bibinfo {pages} {583} (\bibinfo {year}
  {1980})}\BibitemShut {NoStop}%
\bibitem [{\citenamefont {Herbst}\ and\ \citenamefont
  {Wilkins}(1987)}]{HE1987}%
  \BibitemOpen
  \bibfield  {author} {\bibinfo {author} {\bibfnamefont {J.~F.}\ \bibnamefont
  {Herbst}}\ and\ \bibinfo {author} {\bibfnamefont {J.~W.}\ \bibnamefont
  {Wilkins}},\ }\href@noop {} {\emph {\bibinfo {title} {Handbook on the Physics
  and Chemistry of Rare Earths}}}\ (\bibinfo  {publisher} {Elsevier Science},\
  \bibinfo {year} {1987})\BibitemShut {NoStop}%
\bibitem [{\citenamefont {Freeman}\ and\ \citenamefont
  {Watson}(1962)}]{Freeman1962}%
  \BibitemOpen
  \bibfield  {author} {\bibinfo {author} {\bibfnamefont {A.~J.}\ \bibnamefont
  {Freeman}}\ and\ \bibinfo {author} {\bibfnamefont {R.~E.}\ \bibnamefont
  {Watson}},\ }\bibfield  {title} {\bibinfo {title} {Theoretical investigation
  of some magnetic and spectroscopic properties of rare-earth ions},\ }\href
  {https://doi.org/10.1103/PhysRev.127.2058} {\bibfield  {journal} {\bibinfo
  {journal} {Phys. Rev.}\ }\textbf {\bibinfo {volume} {127}},\ \bibinfo {pages}
  {2058} (\bibinfo {year} {1962})}\BibitemShut {NoStop}%
\bibitem [{\citenamefont {Dieke}\ and\ \citenamefont
  {Crosswhite}(1963)}]{Dieke1963}%
  \BibitemOpen
  \bibfield  {author} {\bibinfo {author} {\bibfnamefont {G.~H.}\ \bibnamefont
  {Dieke}}\ and\ \bibinfo {author} {\bibfnamefont {H.~M.}\ \bibnamefont
  {Crosswhite}},\ }\bibfield  {title} {\bibinfo {title} {The spectra of the
  doubly and triply ionized rare earths},\ }\href
  {https://doi.org/10.1364/AO.2.000675} {\bibfield  {journal} {\bibinfo
  {journal} {Appl. Opt.}\ }\textbf {\bibinfo {volume} {2}},\ \bibinfo {pages}
  {675} (\bibinfo {year} {1963})}\BibitemShut {NoStop}%
\bibitem [{\citenamefont {Rau}\ and\ \citenamefont {Gingras}(2018)}]{RA2018}%
  \BibitemOpen
  \bibfield  {author} {\bibinfo {author} {\bibfnamefont {J.~G.}\ \bibnamefont
  {Rau}}\ and\ \bibinfo {author} {\bibfnamefont {M.~J.~P.}\ \bibnamefont
  {Gingras}},\ }\bibfield  {title} {\bibinfo {title} {Frustration and
  anisotropic exchange in ytterbium magnets with edge-shared octahedra},\
  }\href {https://doi.org/10.1103/PhysRevB.98.054408} {\bibfield  {journal}
  {\bibinfo  {journal} {Phys. Rev. B}\ }\textbf {\bibinfo {volume} {98}},\
  \bibinfo {pages} {054408} (\bibinfo {year} {2018})}\BibitemShut {NoStop}%
\bibitem [{\citenamefont {Takahashi}\ and\ \citenamefont
  {Kasuya}(1985)}]{TA1985}%
  \BibitemOpen
  \bibfield  {author} {\bibinfo {author} {\bibfnamefont {H.}~\bibnamefont
  {Takahashi}}\ and\ \bibinfo {author} {\bibfnamefont {T.}~\bibnamefont
  {Kasuya}},\ }\bibfield  {title} {\bibinfo {title} {Anisotropic $p$-$f$ mixing
  mechanism explaining anomalous magnetic properties in {C}e monopnictides.
  {II}. {C}rystal-field splitting in rare-earth pnictides},\ }\href
  {https://doi.org/10.1088/0022-3719/18/13/017} {\bibfield  {journal} {\bibinfo
   {journal} {J. Phys. C: Solid State Phys.}\ }\textbf {\bibinfo {volume}
  {18}},\ \bibinfo {pages} {2709} (\bibinfo {year} {1985})}\BibitemShut
  {NoStop}%
\bibitem [{\citenamefont {Daum}\ \emph {et~al.}(2021)\citenamefont {Daum},
  \citenamefont {Ramanathan}, \citenamefont {Kolesnikov}, \citenamefont
  {Calder}, \citenamefont {Mourigal},\ and\ \citenamefont
  {La~Pierre}}]{PhysRevB.103.L121109}%
  \BibitemOpen
  \bibfield  {author} {\bibinfo {author} {\bibfnamefont {M.~J.}\ \bibnamefont
  {Daum}}, \bibinfo {author} {\bibfnamefont {A.}~\bibnamefont {Ramanathan}},
  \bibinfo {author} {\bibfnamefont {A.~I.}\ \bibnamefont {Kolesnikov}},
  \bibinfo {author} {\bibfnamefont {S.}~\bibnamefont {Calder}}, \bibinfo
  {author} {\bibfnamefont {M.}~\bibnamefont {Mourigal}},\ and\ \bibinfo
  {author} {\bibfnamefont {H.~S.}\ \bibnamefont {La~Pierre}},\ }\bibfield
  {title} {\bibinfo {title} {Collective excitations in the tetravalent
  lanthanide honeycomb antiferromagnet {N}a$_2${P}r{O}$_3$},\ }\href
  {https://doi.org/10.1103/PhysRevB.103.L121109} {\bibfield  {journal}
  {\bibinfo  {journal} {Phys. Rev. B}\ }\textbf {\bibinfo {volume} {103}},\
  \bibinfo {pages} {L121109} (\bibinfo {year} {2021})}\BibitemShut {NoStop}%
\bibitem [{\citenamefont {Ramanathan}\ \emph
  {et~al.}(2023{\natexlab{a}})\citenamefont {Ramanathan}, \citenamefont
  {Kaplan}, \citenamefont {Sergentu}, \citenamefont {Ozerov}, \citenamefont
  {Kolesnikov}, \citenamefont {Minasian}, \citenamefont {Autschbach},
  \citenamefont {Freeland}, \citenamefont {Jiang}, \citenamefont {Mourigal},\
  and\ \citenamefont {La~Pierre}}]{Ramanathan2023A}%
  \BibitemOpen
  \bibfield  {author} {\bibinfo {author} {\bibfnamefont {A.}~\bibnamefont
  {Ramanathan}}, \bibinfo {author} {\bibfnamefont {J.}~\bibnamefont {Kaplan}},
  \bibinfo {author} {\bibfnamefont {J.~A.}\ \bibnamefont {Sergentu},
  \bibfnamefont {D.-C.~Branson}}, \bibinfo {author} {\bibfnamefont
  {M.}~\bibnamefont {Ozerov}}, \bibinfo {author} {\bibfnamefont {A.~I.}\
  \bibnamefont {Kolesnikov}}, \bibinfo {author} {\bibfnamefont {S.~G.}\
  \bibnamefont {Minasian}}, \bibinfo {author} {\bibfnamefont {J.}~\bibnamefont
  {Autschbach}}, \bibinfo {author} {\bibfnamefont {J.~W.}\ \bibnamefont
  {Freeland}}, \bibinfo {author} {\bibfnamefont {Z.}~\bibnamefont {Jiang}},
  \bibinfo {author} {\bibfnamefont {M.}~\bibnamefont {Mourigal}},\ and\
  \bibinfo {author} {\bibfnamefont {H.~S.}\ \bibnamefont {La~Pierre}},\
  }\bibfield  {title} {\bibinfo {title} {Chemical design of electronic and
  magnetic energy scales of tetravalent praseodymium materials},\ }\href
  {https://doi.org/10.1038/s41467-023-38431-7} {\bibfield  {journal} {\bibinfo
  {journal} {Nat. Commun.}\ }\textbf {\bibinfo {volume} {14}},\ \bibinfo
  {pages} {3134} (\bibinfo {year} {2023}{\natexlab{a}})}\BibitemShut {NoStop}%
\bibitem [{\citenamefont {Ramanathan}\ \emph
  {et~al.}(2023{\natexlab{b}})\citenamefont {Ramanathan}, \citenamefont
  {Walter}, \citenamefont {Mourigal},\ and\ \citenamefont
  {La~Pierre}}]{Ramanathan2023B}%
  \BibitemOpen
  \bibfield  {author} {\bibinfo {author} {\bibfnamefont {A.}~\bibnamefont
  {Ramanathan}}, \bibinfo {author} {\bibfnamefont {E.~D.}\ \bibnamefont
  {Walter}}, \bibinfo {author} {\bibfnamefont {M.}~\bibnamefont {Mourigal}},\
  and\ \bibinfo {author} {\bibfnamefont {H.~S.}\ \bibnamefont {La~Pierre}},\
  }\bibfield  {title} {\bibinfo {title} {Increased crystal field drives
  intermediate coupling and minimizes decoherence in tetravalent praseodymium
  qubits},\ }\href {https://doi.org/10.1021/jacs.3c02820} {\bibfield  {journal}
  {\bibinfo  {journal} {J. Am. Chem. Soc.}\ }\textbf {\bibinfo {volume}
  {145}},\ \bibinfo {pages} {17603} (\bibinfo {year}
  {2023}{\natexlab{b}})}\BibitemShut {NoStop}%
\bibitem [{\citenamefont {Wessler}\ \emph {et~al.}(2022)\citenamefont
  {Wessler}, \citenamefont {Roessli}, \citenamefont {Kr\"amer}, \citenamefont
  {Stuhr}, \citenamefont {Wildes}, \citenamefont {Braun},\ and\ \citenamefont
  {Kenzelmann}}]{Wessler2022}%
  \BibitemOpen
  \bibfield  {author} {\bibinfo {author} {\bibfnamefont {C.}~\bibnamefont
  {Wessler}}, \bibinfo {author} {\bibfnamefont {B.}~\bibnamefont {Roessli}},
  \bibinfo {author} {\bibfnamefont {K.~W.}\ \bibnamefont {Kr\"amer}}, \bibinfo
  {author} {\bibfnamefont {U.}~\bibnamefont {Stuhr}}, \bibinfo {author}
  {\bibfnamefont {A.}~\bibnamefont {Wildes}}, \bibinfo {author} {\bibfnamefont
  {H.~B.}\ \bibnamefont {Braun}},\ and\ \bibinfo {author} {\bibfnamefont
  {M.}~\bibnamefont {Kenzelmann}},\ }\bibfield  {title} {\bibinfo {title}
  {Dipolar spin-waves and tunable band gap at the {D}irac points in the 2{D}
  magnet {E}r{B}r$_3$},\ }\href {https://doi.org/10.1038/s42005-022-00965-5}
  {\bibfield  {journal} {\bibinfo  {journal} {Commun. Phys.}\ }\textbf
  {\bibinfo {volume} {5}},\ \bibinfo {pages} {185} (\bibinfo {year}
  {2022})}\BibitemShut {NoStop}%
\bibitem [{\citenamefont {Rau}\ \emph {et~al.}(2014)\citenamefont {Rau},
  \citenamefont {Lee},\ and\ \citenamefont {Kee}}]{PhysRevLett.112.077204}%
  \BibitemOpen
  \bibfield  {author} {\bibinfo {author} {\bibfnamefont {J.~G.}\ \bibnamefont
  {Rau}}, \bibinfo {author} {\bibfnamefont {E.~K.-H.}\ \bibnamefont {Lee}},\
  and\ \bibinfo {author} {\bibfnamefont {H.-Y.}\ \bibnamefont {Kee}},\
  }\bibfield  {title} {\bibinfo {title} {Generic spin model for the honeycomb
  iridates beyond the kitaev limit},\ }\href
  {https://doi.org/10.1103/PhysRevLett.112.077204} {\bibfield  {journal}
  {\bibinfo  {journal} {Phys. Rev. Lett.}\ }\textbf {\bibinfo {volume} {112}},\
  \bibinfo {pages} {077204} (\bibinfo {year} {2014})}\BibitemShut {NoStop}%
\bibitem [{\citenamefont {Rusna\ifmmode~\check{c}\else \v{c}\fi{}ko}\ \emph
  {et~al.}(2019)\citenamefont {Rusna\ifmmode~\check{c}\else \v{c}\fi{}ko},
  \citenamefont {Gotfryd},\ and\ \citenamefont {Chaloupka}}]{RU2019}%
  \BibitemOpen
  \bibfield  {author} {\bibinfo {author} {\bibfnamefont {J.}~\bibnamefont
  {Rusna\ifmmode~\check{c}\else \v{c}\fi{}ko}}, \bibinfo {author}
  {\bibfnamefont {D.}~\bibnamefont {Gotfryd}},\ and\ \bibinfo {author}
  {\bibfnamefont {J.}~\bibnamefont {Chaloupka}},\ }\bibfield  {title} {\bibinfo
  {title} {Kitaev-like honeycomb magnets: Global phase behavior and emergent
  effective models},\ }\href {https://doi.org/10.1103/PhysRevB.99.064425}
  {\bibfield  {journal} {\bibinfo  {journal} {Phys. Rev. B}\ }\textbf {\bibinfo
  {volume} {99}},\ \bibinfo {pages} {064425} (\bibinfo {year}
  {2019})}\BibitemShut {NoStop}%
\bibitem [{\citenamefont {Anisimov}\ \emph {et~al.}(1993)\citenamefont
  {Anisimov}, \citenamefont {Solovyev}, \citenamefont {Korotin}, \citenamefont
  {Czy\ifmmode~\dot{z}\else \.{z}\fi{}yk},\ and\ \citenamefont
  {Sawatzky}}]{AN1993}%
  \BibitemOpen
  \bibfield  {author} {\bibinfo {author} {\bibfnamefont {V.~I.}\ \bibnamefont
  {Anisimov}}, \bibinfo {author} {\bibfnamefont {I.~V.}\ \bibnamefont
  {Solovyev}}, \bibinfo {author} {\bibfnamefont {M.~A.}\ \bibnamefont
  {Korotin}}, \bibinfo {author} {\bibfnamefont {M.~T.}\ \bibnamefont
  {Czy\ifmmode~\dot{z}\else \.{z}\fi{}yk}},\ and\ \bibinfo {author}
  {\bibfnamefont {G.~A.}\ \bibnamefont {Sawatzky}},\ }\bibfield  {title}
  {\bibinfo {title} {Density-functional theory and {N}i{O} photoemission
  spectra},\ }\href {https://doi.org/10.1103/PhysRevB.48.16929} {\bibfield
  {journal} {\bibinfo  {journal} {Phys. Rev. B}\ }\textbf {\bibinfo {volume}
  {48}},\ \bibinfo {pages} {16929} (\bibinfo {year} {1993})}\BibitemShut
  {NoStop}%
\bibitem [{\citenamefont {Anisimov}\ \emph {et~al.}(1997)\citenamefont
  {Anisimov}, \citenamefont {Aryasetiawan},\ and\ \citenamefont
  {Lichtenstein}}]{AN1997}%
  \BibitemOpen
  \bibfield  {author} {\bibinfo {author} {\bibfnamefont {V.~I.}\ \bibnamefont
  {Anisimov}}, \bibinfo {author} {\bibfnamefont {F.}~\bibnamefont
  {Aryasetiawan}},\ and\ \bibinfo {author} {\bibfnamefont {A.~I.}\ \bibnamefont
  {Lichtenstein}},\ }\bibfield  {title} {\bibinfo {title} {First-principles
  calculations of the electronic structure and spectra of strongly correlated
  systems: the {LDA}+${U}$ method},\ }\href
  {https://doi.org/10.1088/0953-8984/9/4/002} {\bibfield  {journal} {\bibinfo
  {journal} {J. Phys.: Condens. Matter}\ }\textbf {\bibinfo {volume} {9}},\
  \bibinfo {pages} {767} (\bibinfo {year} {1997})}\BibitemShut {NoStop}%
\bibitem [{\citenamefont {Stevens}(1952)}]{Stevens1952}%
  \BibitemOpen
  \bibfield  {author} {\bibinfo {author} {\bibfnamefont {K.~W.~H.}\
  \bibnamefont {Stevens}},\ }\bibfield  {title} {\bibinfo {title} {Matrix
  elements and operator equivalents connected with the magnetic properties of
  rare earth ions},\ }\href {https://doi.org/10.1088/0370-1298/65/3/308}
  {\bibfield  {journal} {\bibinfo  {journal} {Proc. Phys. Soc. A}\ }\textbf
  {\bibinfo {volume} {65}},\ \bibinfo {pages} {209} (\bibinfo {year}
  {1952})}\BibitemShut {NoStop}%
\end{thebibliography}%

\end{document}